\documentclass[a4paper,twocolumn,11pt,accepted=2021-06-18]{quantumarticle}
\pdfoutput=1
\usepackage[utf8]{inputenc}
\usepackage[english]{babel}
\usepackage[T1]{fontenc}
\usepackage{amsmath}
\usepackage{amssymb}
\usepackage{hyperref}

\usepackage{tikz}
\usepackage{lipsum}
\usepackage{graphicx}
\usepackage{qcircuit}
\usepackage{physics}

\definecolor{LightGrey}{gray}{0.85}
\definecolor{LightGreen}{rgb}{0.60,1,0.2}
\definecolor{UberLightGreen}{rgb}{0.80,1,0.6}
\definecolor{LightOrange}{rgb}{1,1,0.39}

\newcommand{\be}{\begin{equation}}
\newcommand{\ee}{\end{equation}}

\newcommand{\mr}{\ensuremath{\mathrm}}

\newlength\figureheight
\newlength\figurewidth

\def\ket#1{\left|#1\right>}
\def\bra#1{\left<#1\right|}

 \def\ee{\mathord{\rm e}}

\def\ii{\mathord{\rm i}}

\def\half{\textstyle\frac{1}{2}}

\renewcommand{\ii}{{\rm i}}
\renewcommand{\ee}{{\rm e}}

\definecolor{mygrey}{gray}{0.94}
\definecolor{myred}{rgb}{1,0.,0.3}

\newcommand{\id}{\mathbb{I}}

\usepackage[numbers,sort&compress]{natbib}
\begin{document}

\title{Crosstalk Suppression for Fault-tolerant Quantum Error Correction with Trapped Ions}

\author{Pedro Parrado-Rodr\'iguez}
\affiliation{%
Department of Physics, College of Science, Swansea University, Singleton Park, Swansea SA2 8PP, United Kingdom}%
\orcid{0000-0002-6201-2723}

\author{Ciar\'{a}n Ryan-Anderson}%
 \affiliation{%
Department of Physics, College of Science, Swansea University, Singleton Park, Swansea SA2 8PP, United Kingdom}%
\affiliation{Honeywell Quantum Solutions, 303 S. Technology Ct., Broomfield, Colorado 80021, USA}

\author{Alejandro Bermudez}%
\affiliation{Departamento de F\'isica Te\'orica, Universidad Complutense, 28040 Madrid, Spain.}%

\author{Markus M\"uller}%
 \affiliation{Institute for Theoretical Nanoelectronics (PGI-2), Forschungszentrum Jülich, 52428 Jülich, Germany}
 \affiliation{JARA-Institute for Quantum Information, RWTH Aachen University, 52056 Aachen,
Germany}


\maketitle

\begin{abstract}
Physical qubits in experimental quantum information processors are inevitably exposed to different sources of noise and imperfections, which lead to errors that typically accumulate hindering our ability to perform long computations reliably. Progress towards scalable and robust quantum computation relies on exploiting quantum error correction (QEC) to actively battle these undesired effects. In this work, we present a comprehensive study of crosstalk errors in     a quantum-computing architecture based on a single string of ions confined by a    radio-frequency  trap,     and manipulated by individually-addressed laser beams. This type of errors 
affects spectator qubits that, ideally, should remain unaltered during the application of  single- and two-qubit quantum gates    addressed at a different set of active qubits. We microscopically model crosstalk errors from first principles and present a detailed study showing the importance of using a coherent vs incoherent error modelling and, moreover, discuss strategies to actively suppress this crosstalk at the gate level. Finally, we study the impact of residual crosstalk errors on the performance of fault-tolerant QEC numerically, identifying the experimental target values that need to be achieved in near-term trapped-ion experiments to reach the break-even point for beneficial QEC with low-distance topological codes.
\end{abstract}

\section{\bf \label{sec:intro} Introduction}
Quantum computation aims at manipulating delicate entangled states to achieve functionalities beyond those presented by classical devices. Robust large-scale quantum computers will likely require quantum error-correction (QEC) to exploit the wide range of applications offered by a universal quantum processor~\cite{nielsen-book}. In particular, scalable quantum error correction codes (QECCs) preserve quantum information by encoding it redundantly in a set of physical qubits~\cite{nielsen-book} such that, in principle, arbitrary levels of protection can be achieved by increasing the number of redundant physical qubits while employing active detection and correction of errors, provided physical noise rates lie below the critical threshold of the corresponding QECC. Together with a fault-tolerant methodology \cite{shor1996faulttolerant}, which forbids the uncontrolled proliferation of errors by the specific design of the scalable QEC circuits, this yields one of the most promising approaches towards large-scale quantum computation. However, to achieve the required levels of protection,  there are experimental and theoretical challenges that need to be addressed.  In particular, identifying and mitigating noise sources so that minimal QECCs are shown to outperform their physical/bare counterparts is considered a break-even point in the road-map for realising QEC for large-scale quantum computers \cite{Devoret2013,Bruzewicz2019}.

In several qubit-based quantum processors \cite{s-ladd-nature-464-45}, the implementation of some of the building blocks of QEC, such as the encoding of information in a QECC and the detection and correction of errors without altering the encoded information, has been demonstrated, for example, using trapped ions \cite{nigg-science-345-302,PhysRevX.6.031030, Bermudez2017,SchindlerScience1059,kielpinski2001recent,Chiaverini2004,Olsacher2020, Negnevitsky2018,egan2020faulttolerant,Zhu2019}, superconducting circuits \cite{Kjaergaard_2020, Kelly2015,PhysRevLett.117.210505, ofek2016demonstrating, Andersen2020}, nuclear magnetic resonance \cite{s-zhang-prl-109-100503, s-knill-prl-86-5811}, or nitrogen-vacancy centres \cite{Waldherr_2014,Unden2016}. Recently, and in parallel to these efforts, there has also been much progress in protecting quantum information in oscillator-based bosonic QEC encodings (see, e.g.~\cite{Terhal_2020,Noh_2020} using trapped ions \cite{Fl_hmann_2019,deneeve2020error,Fluhmann2018} and superconducting setups \cite{Campagne-Ibarcq2020,Ofek2016}).

In this study, we will focus on qubit-based QEC with trapped ions. To date, trapped-ion experiments have shown high fidelity  gates and long coherence times \cite{Erhard_2019,Langer2006HighFQ,Ballance_2016,hartyFidelity,gaeblerFidelity,Wang_2017, Schafer2018}, making them one of the most promising candidates for the future realisation  of large-scale quantum computers. There are currently two approaches for the scalability of trapped-ion  systems for QEC. The first approach is the single-string ion trap, where all the ions are placed in    modules consisting of a single, linear    radio-frequency trap, and coupled by photonic interconnects \cite{PhysRevA.89.022317, Nickerson2014}. On the other hand, there is a so-called shuttling-based approach to scalability~\cite{Kielpinski2002},  which utilises microfabricated  traps. In this approach, the ions are moved between storage and manipulation zones, where they are subjected to single-qubit gates and fluorescence measurements as needed or otherwise  merged with other ions to implement  entangling   gates or sympathetic laser cooling. After applying the desired sequence of operations, the ions can be split and    shuttled back to the storage zones. There they remain as spectators of the operations on other active ions in the manipulation zones~\cite{Bermudez2017, kaushal2019shuttlingbased, kielpinski2001recent, Kielpinski2002,Pino2020, Kaufmann2017,Wan2019}. Our study is motivated by new experimental capabilities that have recently emerged in the single-string trapped-ion modules. These new experimental capabilities include the possibility to perform high-fidelity  entangling gates addressed on specific subsets of ions \cite{Landsman_2019,aodthesis,privatecomm} , leading to effective all-to-all connectivity for two-qubit entangling gates. These capabilities have motivated the study of the single-string ion trap scenario as a viable approach to implementing an operational logical qubit \cite{Linke_2017,  Bermudez2017}. Informed by these results, we focus our study on the single string ion traps, which can eventually be combined with the shuttling-based or photonic-interconnect techniques for further scalability. 

Topological codes \cite{Kitaev_2003, Kitaev1997,bombin-prl-97-180501,Lidar2012,Terhal_2015}, a specific class of scalable QECCs, are currently being studied as one of the main options for future  QEC as they have high error thresholds. Moreover, they are designed such that the physical qubits only need to be manipulated locally, i.e. requiring interactions between neighbours. Local errors that stem from such manipulations can then be efficiently detected and corrected by the QECC. To achieve this scalable protection, the  lattices where the physical qubits reside must be enlarged, and the fidelity of the operations on the physical qubits needs to be reduced below the aforementioned threshold. Some experiments have already demonstrated levels of control on the required regime for small systems (e.g.~ \cite{Ballance_2016}),  but scaling up  the experiments while maintaining the  desired degree of precision remains a challenge. To tackle this, various low-distance QEC codes at the reach of different state-of-the-art technologies, are being actively investigated. For such low-distance codes, it is of primary importance to use fault-tolerant (FT) circuit designs so that errors do not proliferate due to unsuitable circuit layouts, allowing to exploit the full correcting power of the QECC.  Such fault-tolerant circuit constructions will serve the purpose of reaching the break-even point of the advantage of QEC in small devices and provide crucial information that can guide future strategies in progress towards large-scale quantum computers. The 7-qubit colour code~\cite{bombin-prl-97-180501}, also known as the Steane code ~\cite{steane-prl-77-793}, is one of the workhorses for such low-distance QECCs, and various techniques for an FT syndrome readout have been devised~\cite{shor_ft_qec,aliferis_ft_qec}. Due to present experimental capabilities, there has been intense recent activity developing resource-efficient FT schemes, such as FT readout techniques that exploit a single ancilla for certain types of microscopic noise~\cite{singleflag-pra-brown, Chamberland_2018_q, Chamberland_2020_1, Chamberland_2020_2}. Another example is using the so-called flag-based readout \cite{Chao2018,Lidar2012,chaoReichardt,Reichardt-new}, which substantially reduces the number of required  ancillary qubits,  encompassing noise models beyond the single-qubit limit. These FT constructions exploit the full correcting power of  the 7-qubit colour code, which allows for the correction of any single  fault in the encoded logical qubit. In turn, this yields a different scaling of the logical error rate to the physical error rate, which allows for the existence of  pseudo-thresholds and is crucial for the demonstration of the advantage of QEC with small-scale devices. Colour codes distinguish themselves from other existing QECCs because of their capability of a transversal implementation of the entire group of Clifford gates \cite{bombin-prl-97-180501}. 

Another important aspect for the progress of  QEC is to inform the theoretical descriptions, which typically assess a particular QECC under a specific noise model, with experimental results such that the noise model can be made as realistic as possible. In the context of trapped ions, a recent effort along these lines has been accomplished for the shuttling-based approach~\cite{Bermudez2017,Trout_2018,PhysRevA.99.022330, 1810.09199} as well as for the single-string approach~\cite{PhysRevA.100.032325,Majumder2020,Brown:2019}. In this manuscript, motivated by the recent developments of new addressing optics \cite{Landsman_2019, privatecomm}, which allow addressing subsets of ions in strings, we focus on the single-string approach. Specifically, we present a microscopically-motivated error model that accounts for imperfections in all the required operations. A particular focus is placed on crosstalk errors. As described in \cite{sarovar2019detecting},  crosstalk errors cause undesired dynamics that violate either
(or both) of two principles: spatial locality (i.e. gates should only affect the target qubits) and independence  of local operations (i.e. the effect of a gate should be independent of other gates being applied to the system). In ion traps, crosstalk is caused by residual illumination of laser light on neighbouring ions when applying single- and two-qubit gates. This error source can be  effectively minimised for single-qubit gates \cite{nigg-science-345-302} using composite pulse sequences and dynamical decoupling techniques \cite{2014dynamicalReview} but becomes a more delicate issue when arising in the context of two-qubit entangling gates. 
With the use of carefully constructed pulse sequences, it is possible to suppress unwanted couplings on multi-qubit coupling operations, as shown in  \cite{Leung_2000,Hayes_2014}. 
In this work, we consider two different models of crosstalk errors caused by this residual illumination. In the context of fault-tolerant quantum computation (FTQC),  fault-tolerant constructions assume that errors only happen on those qubits that are involved in the gates. Thus, crosstalk represents an error source that potentially breaks the fault-tolerant character of FT circuit constructions, as it can induce error processes affecting groups of physical qubits that would, otherwise, not interact through the gates. In this way, the  FT design of the circuits is compromised, possibly leading to the proliferation of errors. Therefore, if crosstalk is not adequately accounted for, or suppressed, it can have a damaging effect on the performance of the QEC protocol and break fault tolerance \cite{Piltz2014, Murali_2020,Vandersypen_2005,Reagor_2018, Harper_2020, Aude_Craik_2017}.

Generally, one can attempt to mitigate errors in the whole computation at different levels, such as at the level of gates \cite{Vandersypen_2005,Viola_1999,Goerz2015,Mabuchi2005,RajControl,wiseman_milburn_2009}, circuits  \cite{Chao2018,chaoReichardt, Reichardt-new}, full QECCs \cite{Lidar2012}, or even entire algorithms \cite{Murali_2020, 2014dynamicalReview}. Recent work \cite{Brown:2019} has described how to combat crosstalk in ion traps on the level of QECCs and circuits by choosing the arrangement of qubits in the QECC circuitry to minimise the fault-tolerant breaking effects of crosstalk, and by comparing the performance of different QECCs in the compass-code QEC family. As a complementary approach, we will investigate a technique that suppresses crosstalk  directly at the gate level. Gate-level mitigation of noise can suppress and change the nature of any remaining effective noise. Thus, one may want to consider such techniques when evaluating the performance and merits of different QEC protocols. In particular, we will concentrate on a flag-based scheme~\cite{Reichardt-new} for the distance-three seven-qubit colour code. Our purpose is to analyse whether the   refocussing technique, to be described in detail below, allows this QEC-operated logical qubit to outperform a physical qubit  in the presence of microscopic errors that include crosstalk. Here, we want to identify the specific regimes where this QEC advantage becomes possible, informing future hardware improvements. We will also explore the consequences of the coherent nature of crosstalk and the extent to which it is possible to suppress its impact on QEC. We consider near-term noise rates and crosstalk, building on previous works \cite{1810.09199}, and extend the noise models towards a more complete description. It should be noted that although our study focuses on a trapped-ion quantum processor and the example of the 7-qubit colour code, the error mitigation technique and analysis discussed here can be adapted to other quantum processor architectures, or other QECCs and algorithms.

The organisation of this paper is as follows. In Section~\ref{sec.background}, we present the ion-trap operation toolbox, examine the error model for the ion trap, and review the QEC protocol we will be focusing on. Most importantly, we introduce  two error models of crosstalk used throughout the remainder of the paper. In Section~\ref{sec.problem}, we discuss how unmitigated crosstalk affects the performance of the QEC protocol. In Section~\ref{sec.results}, we describe the  refocussing technique intended to suppress crosstalk and derive the form and strength of residual crosstalk noise after applying the  refocussing technique. We then examine how  refocussing improves the performance of the QEC protocol by extensive numerical simulations.  Finally, in Section~\ref{sec.conclusion}, we briefly discuss further avenues for crosstalk mitigation and present our conclusions.


\section{\bf Trapped-ion toolbox for QEC \label{sec.background}}

\subsection{Static ion-string  gate set and noise model\label{sec.back.ions}\label{sec.noise}}

We focus on the single-string ion trap approach and a single ion species, allowing one to encode qubits in an optical transition that can be operated with addressed laser light in the visible spectrum. For example, in  $^{40}{\rm Ca}^+$ ions, the $\ket{0}$ state corresponds to the ground state $\ket{S_{1/2},m_j=-1/2}$, and the $\ket{1}$ state to the metastable excited state $\ket{D_{5/2},m_j=-1/2}$. We now briefly catalogue the set of physical gates we will utilise for these optical qubits in a static-string ion trap to implement QEC. References~\cite{1810.09199, Schindler2013} provide a more detailed review of these gates.

\subsubsection{Single and two-qubit gates} 

For single-qubit gates, ions in the string can be individually addressed with laser beams  \cite{singleqaddressing} to apply a rotation

\begin{align}
\label{eq:single_qubit_roattion}
    P_i(\theta) = e ^{-i  \frac{\theta}{2} P_i},
\end{align}

\noindent where $P \in \{X, Y, Z\}$    belongs to the set of Pauli matrices, and $\theta$    is the angle of rotation. The errors in the single-qubit gates stem from fluctuations in the laser beam's intensity and phase, which lead to over/under-rotations and changes in the axis of rotation, respectively~\cite{PhysRevA.99.022330}. In the present work, we model these errors  by a symmetric depolarising channel, which admits a decomposition in terms of Kraus operators~\cite{nielsen-book}, namely

\begin{align}
\label{kraus_decomp}
   \rho\to D(\rho)=\sum_nK_n\rho K_n^\dagger. 
\end{align}

For the depolarising channel, the Kraus operators are $K_n\in\{\sqrt{1-p_{\rm 1q}}\,\,\id,\sqrt{p_{\rm 1q}/3}\,X,\sqrt{p_{\rm 1q}/3}\,Y,\sqrt{p_{\rm 1q}/3}\,Z\}$, where we have introduced the  error rate per single-qubit gate $p_{\rm 1q}$, the value of which  depends on the sources of gate imperfections listed above, and will be fixed to realistic values as achieved in current experiments, and as detailed below.

\begin{figure}[!ht]
\begin{centering}
\includegraphics[width=0.99\columnwidth]{ 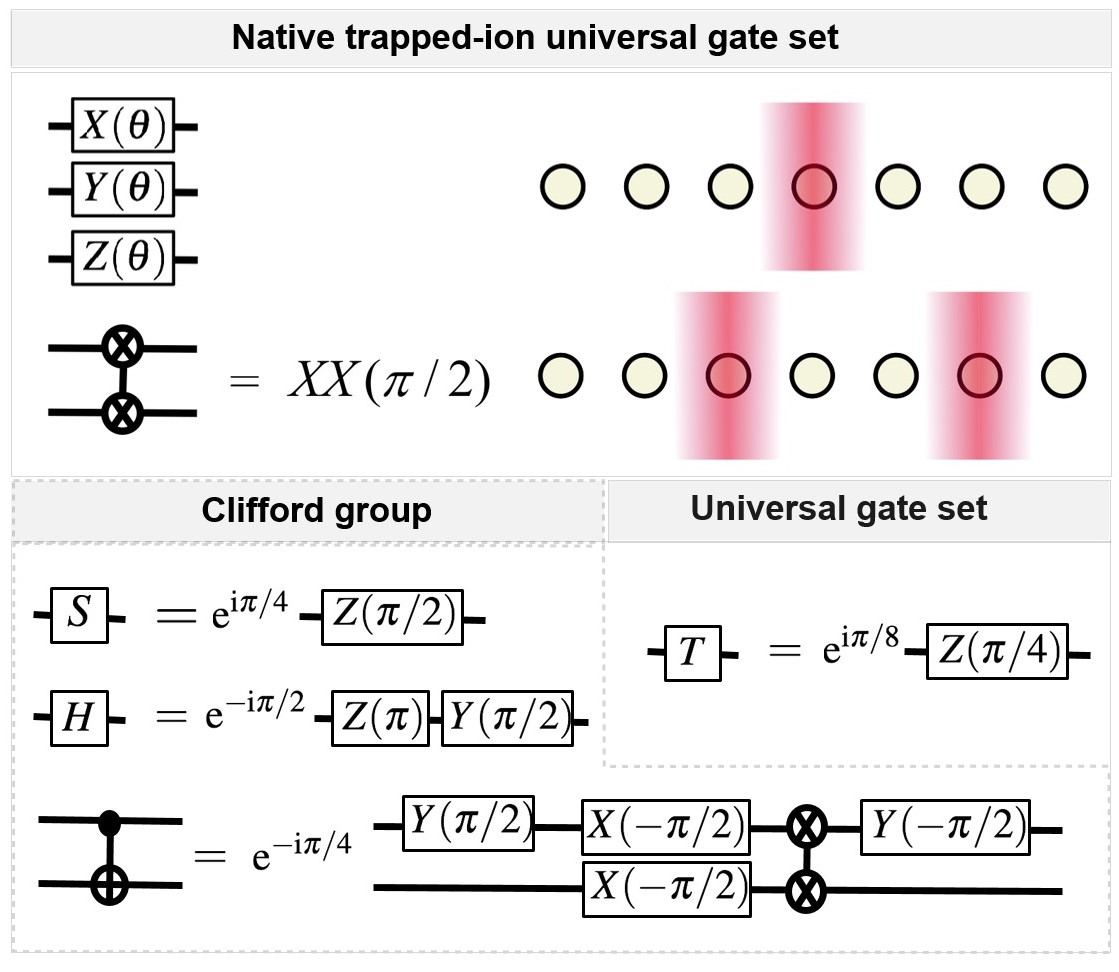}
 \caption{\textbf{Gate set for the static ion-string approach:} By using addressed single-qubit rotations and the two-qubit entangling MS gate, we have a universal gate set available. The gate compilation of the CNOT gate using the MS gate and single-qubit rotations is not unique. For other useful gate compilations, see \cite{Maslov2017}.}
\label{fig:gatesetfig}
\end{centering}
\end{figure}

For multi-qubit entangling operations, a popular and widely used two-qubit entangling gate for optical qubits is known as the Mølmer–Sørensen (MS) gate \cite{molmer-prl-82-1835,Sorensen1999,Roos2008}. This gate equips the static string with complete qubit connectivity, as it exploits a common vibrational mode as a quantum data bus, which can  mediate the entangling gate operation between an arbitrary pair (or larger groups) of qubits that share this common mode, provided that the laser beams  uniformly illuminate this pair (or group) and do not illuminate the remaining ions.  
 For this work, we concern ourselves only with the MS gate in the $XX$-basis which, when addressed to  the pair of ions with indexes $i,j$ within the static string, can be expressed through the following unitary 
\begin{align}
\label{eq:MS_gate}
    XX_{ij}(\theta) = \exp \left( -i  \frac{\theta}{2} X_iX_j \right),
\end{align}

where $ X_iX_j$ is a tensor product of Pauli $X$-type operators on qubits $i$ and $j$,    and $\theta$ is the MS-gate angle. 
We focus on this version of the MS gate because the fully-entangling MS gate $XX_{ij}\left(\pi /2\right)$, together with single-qubit rotations, are sufficient to enact the CNOT gate (see Fig.~\ref{fig:gatesetfig}), which is the two-qubit gate most often used to describe QEC   circuitry \cite{Nebendahl2015,Maslov2017}. 

These gates allow the implementation of the universal set of unitaries $\{ H, S, T, CNOT\}$ \cite{Boykin:1999}  by using combinations of the MS gate and single-qubit rotations (see Fig.\ref{fig:gatesetfig} for more details). Furthermore, for the static-string approach with single-ion laser addressing, the system has full two-qubit connectivity with this gate set. This allows for a better connected system of qubits, which benefits the performance, reducing the number of gates required to reproduce a general circuit \cite{Linke_2017}. 

The error model for the MS gate finds its sources in the  residual spin-phonon entanglement with the different levels of oscillation of the ion chain, as well as fluctuations on the intensity and phase of the laser beam and the global magnetic fields - see \cite{Ballance_2016} for a recent comprehensive analysis. An effective error model was derived in \cite{1810.09199} from microscopic sources and consists of the stochastic application of Pauli-type errors. The details of this error model can be found in Appendix \ref{sec:appendix_errormodels}.
  
Additionally, as discussed below, the MS gate error model includes crosstalk between the gate ions and the neighbouring ions.

\subsubsection{Crosstalk errors} 

In general, crosstalk  noise arises from any gate that unintentionally affects qubits that should not be involved and should thus remain mere spectators. Operational crosstalk in optically-manipulated trapped-ion systems stems from imperfect addressing, such that light      intended to be addressed to a specific set of active qubits also illuminates some spectator qubits, which are the qubits 
 arranged in the immediate neighbourhood of the active qubits. The origin of this imperfect addressing is generally given by the finite size of the laser focusing spot at the  position of the ions. In this work, we consider two different models for crosstalk errors in the MS gate: {\it the entangling crosstalk model} and {\it the Stark-shift crosstalk model}. It should be noted that whereas single-qubit gates also generate crosstalk, this noise can be greatly suppressed by techniques such as composite sequences \cite{Merrilladdressing} or dynamical decoupling \cite{Viola_1999} (see Appendix~\ref{sec:appendix_A} for a further discussion). The qubit measurement can also lead to  crosstalk, as the measurement process also leads to the scattering of photons from the measured ions, which can subsequently interact with the remaining spectator ions in the string. We note that it is possible to prevent this type of crosstalk  by applying spectroscopic decoupling pulses to nearby spectator ions~\cite{nigg-science-345-302}, such that the scattered photons become off-resonant and the crosstalk is minimised (see Appendix ~\ref{appendix_measurement}). Therefore, in the following, we will only focus on crosstalk in MS-gates.   
 
\begin{figure}[h!]
\begin{centering}

\includegraphics[width=0.95\columnwidth]{ 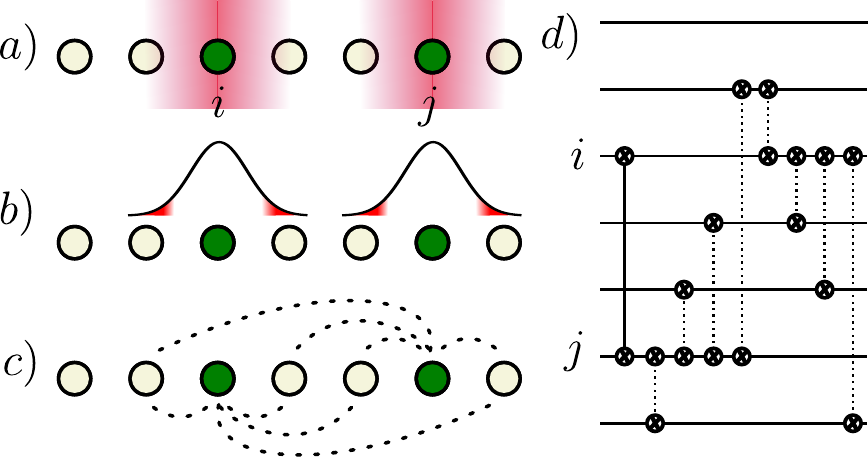}
\\
\caption{\textbf{Addressed entangling gates in the static ion string approach:} \textbf{a)} Light is focused on the ions $i$ and $j$ encoding the qubits involved in the 2-qubit gate. \textbf{b)} Residual light on `neighbour' ions in the direct spatial vicinity of the target `gate' ions causes small residual crosstalk between the gate and neighbouring spectator ions. \textbf{c)} These interactions result in residual effective two-body interactions between each target (gate) ion and each of the neighbouring spectator ions (indicated by dashed lines). The form and quantitative dependence of these crosstalk terms are discussed in the main text. \textbf{d)} Circuit representation of the MS gate (solid line) and crosstalk events (dashed lines) resulting from the gate using the entangling error model for crosstalk, as described in the text. }

\label{fig.Gate-and-neighbor-ions}
\end{centering}
\end{figure}

In the entangling crosstalk model, the leading order effect of crosstalk, as deduced by our microscopic analysis (see Appendix~\ref{sec:appendix_A}), is the partial application of MS operations between the set of gate ions and the set of neighbouring ions (Fig.~\ref{fig.Gate-and-neighbor-ions}). For an MS-gate rotation of an arbitrary angle $\theta$, the leading order noise can be expressed as a unitary that involves active qubits and neighbouring spectators of the following form 

\begin{equation}
E(\theta)_{\text{crosstalk}} = \prod_{\substack{g \in G,\\ N \in \text{neig}(G)}} \exp\left(-i \; \epsilon_{CT}\frac{\theta}{2}  X_g X_N\right).
\label{eqCrosstalk}
\end{equation}
\noindent 

where $G = \{i,j\}$ is the set of gate qubits the ideal MS gate acts on,  $\text{neig}(G)$ is the set of qubits that are neighbours to $i$ and/or $j$ but do not include $i$ or $j$, and $\epsilon_{CT} =\Omega_n/\Omega$ depends on the fraction of the  light intensity illuminating the spectator qubits with respect to the intensity addressed to the active qubits. We do not consider any effects of crosstalk on next to nearest neighbours, as we assume a sufficiently narrow Gaussian profile of the laser light. The amplitude of the electric field away from the ions becomes exponentially suppressed, and therefore we neglect the effects of crosstalk beyond nearest neighbours. In particular, $\Omega$ is the Rabi frequency for the two-ion MS gate and $ \Omega_n $  is the relative Rabi frequency of the residual light on the neighbouring ions. For a single gate-neighbour crosstalk pair of a gate qubit $g$ and a neighbouring qubit $n$, the error channel is $ \rho\to D_{\rm 2q}^{\rm ct}(\rho)$, where

\begin{align}
     D_{\rm 2q}^{\rm ct}(\rho)&= \cos^2\left(\tfrac{ \theta\epsilon_{CT} }{2}\right)  \; \rho \nonumber\\
&    +\sin^2\left(\tfrac{ \theta\epsilon_{CT} }{2}\right) \; X_g X_n \; \rho \; X_g X_n    \nonumber\\
    &+ i \frac{1}{2}  \sin\left( \theta\epsilon_n \right) \; [\rho, \; X_gX_n]. 
    \label{eq.coherent.crosstalk}
\end{align}

\noindent In this expression, we can identify the probability of a single crosstalk-error event as $ p_c = \sin^2(\epsilon_{CT} \, \theta /2)$,    which leads to a correlated 2-qubit bit-flip error on one of the active qubits and a neighbouring spectator qubit. It is clear from this perspective that the entangling crosstalk error model should be carefully considered in the context of the FT circuit design, as a single MS-gate error can turn into a dangerous pair of errors if  both the active and the spectator ions belong to the set of physical data qubits.

We can use  this probability $p_c$ to define a stochastic/incoherent version of the crosstalk channel described in Eq.~\eqref{eq.coherent.crosstalk}. That is, the Kraus decomposition of Eq.~(\ref{kraus_decomp}) for incoherent crosstalk is

\begin{equation}
\label{eq:Kraus_incoherent}
\begin{split}
    K_{c,0} &= \sqrt{1-p_c} \; \id , \\
    K_{c,1} &= \sqrt{p_c} \; X_gX_n. 
\end{split}
\end{equation}

\noindent When an ion lies in between two gate ions, it will receive twice the residual light from the gate. For that case, the probability of crosstalk is effectively increased by a factor of 4.  Note that we model both incoherent and coherent crosstalk, as we will compare both models in the simulations. The coherent nature of crosstalk can have important implications for QEC performance. On the one hand, it can lead to a coherent build-up of  errors. On the other hand, the coherent nature also offers the possibility to greatly suppress crosstalk using  refocussing schemes, as discussed in the following sections. \\

The second crosstalk model we consider is associated with an alternative way of implementing addressed entangling MS gates. Here, the idea is to apply a power imbalance between the intensities of the bichromatic light fields coupling to the red and blue phonon sidebands, respectively. This difference in intensities leads to an ac-Stark shift, which will differ for active and spectator ions. By tuning the power imbalance appropriately, the gate qubits will be subjected to a  two-photon resonance and undergo the ideal MS gate dynamics, as desired, in contrast to the spectator ions that will effectively be exposed to weak and far off-resonant light. As a consequence, this setting avoids entangling interactions of spectator qubits with the active ions. The leading-order effect of the residual crosstalk is an ac-Stark shift on the spectator ions, resulting in small unwanted, but systematic $Z$-rotations, which can be accounted for by 

\begin{equation}
E^{(2)}_{\text{crosstalk}} = \prod_{ n \in \text{neig}(G)} \exp\left(-i \; \mu_n \frac{\pi}{4} Z_n\right),
\label{eq:crosstalk-starkshift-coherent}
\end{equation}

\noindent where we use the same notations as in Eq.~\eqref{eqCrosstalk} and introduce $\mu_n$ as the residual ac-Stark shift from the off-resonant sidebands that act on the spectator ions due to crosstalk. A detailed derivation of this error model can be found in Appendix \ref{sec.appendixACstarkshift}.

An incoherent approximation of this second crosstalk noise model, suitable to be used in stabiliser-based numerical simulations, is given by a single-qubit dephasing channel with the following Kraus decomposition~\eqref{kraus_decomp} on each of the neighbouring spectator qubits

\begin{equation}
\begin{split}
    K_{c,0} &= \sqrt{1-p_c} \; \id , \\
    K_{c,1} &= \sqrt{p_c} \; Z_n, 
    \end{split}
    \label{eq:Kraus_incoherent_ac}
\end{equation}

with $p_c = \sin^2(\mu_n \pi/4)$.  

\subsubsection{Errors on idling qubits}
We consider here trapped-ion qubits encoded in an optical qubit transition, as e.g.~in $^{40}{\rm Ca}^+$ qubit. When idling, these qubits suffer three types of noise: dephasing, amplitude damping, and leakage. Dephasing, which arises, e.g. from ambient fluctuating magnetic fields or laser frequency drifts, will be modelled as temporally and spatially uncorrelated in this work. Amplitude damping results from spontaneous decay from the metastable state $\ket{1}$ to the ground state $\ket{0}$. Leakage occurs if the state $\ket{1}$ decays instead to the Zeeman sublevel $\ket{\ell}= \ket{S_{1/2},m_j=+1/2}$, which lies outside of the qubit subspace, as opposed to the ground state $\ket{0} = \ket{S_{1/2},m_j=-1/2}$. For the experimental system we study, $T_1=  1/\Gamma$ is around 1.1 seconds, and the fraction of leakage with respect to amplitude damping is $\Gamma_l = \frac{4}{9}\Gamma_d$ \cite{1810.09199}, with $\Gamma=\Gamma_{d}+\Gamma_l$. While one can fully simulate the amplitude-damping/leakage channel using the Kraus operators and circuit identities, as discussed in Ref.~\cite{1810.09199}, for our simulations we approximate both channels using a Clifford approximation, similar to the model presented in Ref.~\cite{Gutierrez:2013}, which can be efficiently implemented in stabiliser simulations. 

To numerically simulate this combination of error channels, we first check that the state is not in $\ket{0}$ or $\ket{\ell}$ already. Then, we use a random number to sample the probability distribution of a decay event  with $p = 1-\exp(-\Gamma \Delta t)$,  where $\Delta t$ denotes the time step, and a second random number, according to the electronic branching ratio, determines if the qubit decays to the leaked state $\ket{\ell}$ or to the ground state $\ket{0}$. The corresponding Kraus operators~\eqref{kraus_decomp} for our approximation are

\begin{equation}\label{eq.leakageKraus}
\begin{split}
    K_0 &= \sqrt{1-p} \,\id,  \hspace{1ex} K_2 = \sqrt{p} \ket{0}\bra{0}, \hspace{1ex}   K_3 = \sqrt{p} \ket{\ell}\bra{\ell}\\
    K_4 &=  \sqrt{p\tfrac{\Gamma_d}{\Gamma}} \ket{0}\bra{1}, \hspace{1ex}  K_5=  \sqrt{p\tfrac{\Gamma_l}{\Gamma}} \ket{\ell}\bra{1}.
    \end{split}
    \end{equation}

While the physical origin of leakage in qubits encoded in an optical transition can be similar to amplitude damping, special treatment needs to be made of leaked states as they are no longer in the computation basis. For leaked states, we follow a similar procedure as outlined in Ref.~\cite{1308.6642}. In the simulation, if a qubit leaks, it is projected to the $\ket{0}$ state and a classical variable is used to record that the qubit has leaked. If leaked qubits are exposed to laser fields that aim at realising single-qubit unitaries, they do not evolve as the laser fields are off-resonant and do not bring them back to the computational qubit subspace. If a leaked qubit would take part in an $MS$ gate, neither the $MS$ gate nor the noise associated with the $MS$ is applied. Furthermore, as discussed in~\cite{1810.09199}, when a non-leaked qubit takes part in a $MS$ with a leaked qubit, no additional noise is applied to the non-leaked qubit. Measurement and state preparation operations reset leaked qubits to the $\ket{0}$ state since both operations are followed or realised by optical pumping into $\ket{0}$. Thus, during measurement and state preparation, the value of the classical variable used to track leaked qubits is modified to indicate that qubits undergoing these operations are no longer leaked. \\

By a specifically tailored leakage repumping protocol \cite{1810.09199} we can reset a leaked state reset to $\ket{0}$ while leaving computational states largely unaffected. However, this process can be itself faulty and lead to leakage, dephasing and amplitude damping errors, as discussed in detail in Appendix \ref{sec:appendix_errormodels}. In the QEC protocols we study, we apply this leakage repumping sequence before each stabiliser measurement round. 

\subsubsection{State initialisation and measurement errors} 

As for state preparation and measurement, ions are prepared in the state $\ket{0}$ and measured in the $Z$-basis. Ions  may additionally be re-prepared in the $\ket{0}$ state at any arbitrary point during computation. Faults on these gates are modelled by applying Pauli X errors after state preparation gates and before the measurement gates. 
Additionally, we include that imperfections in the preparation of the $\ket{0}$ state can lead to a leakage error.

\begin{table}
  \centering
  \resizebox{0.95\columnwidth}{!}{%
  \begin{tabular}{|l |c|c|} \hline
    Operation & Anticipated & Anticipated  \\
      &   duration &   Infidelity \\\hline
      Two-qubit MS gate       & 15$\mu$s & $2 \cdot 10^{-4}$ \\ \hline
     One-qubit gate           & 1$\mu$s & $1 \cdot 10^{-5}$ \\ \hline
     Measurement              & 30$\mu$s & $1 \cdot 10^{-4}$ \\ \hline
     Qubit reset              & 10$\mu$s &$1 \cdot 10^{-4}$ \\ \hline
     Re-cooling               & 100$\mu$s & $\bar{n} < 0.1$ \\ \hline
     Leakage repumping                 & 20$\mu$s & $2 \cdot 10^{-4}$ \\ \hline 
  \end{tabular}
  }
  \caption{{\bf Extended trapped-ion QEC toolbox}. Description of
    expected near-term experimental trapped-ion capabilities for a QCCD approach to
    FT QEC (Values taken from Ref.~\cite{1810.09199}). These values correspond to a particular ion trap setup. Currently, better values for particular gates have already been achieved in  dedicated experiments. For a more detailed review of the current best fidelities and durations, the reader can consult Table 1 in  \cite{Bruzewicz2019}.}
  \label{tab:summary_toolbox_y3}

\end{table}

\subsection{Flag-based QEC with colour codes}\label{sec:flagscheme}

\begin{figure}[!ht]
\begin{centering}
\includegraphics[width=0.95\columnwidth]{ 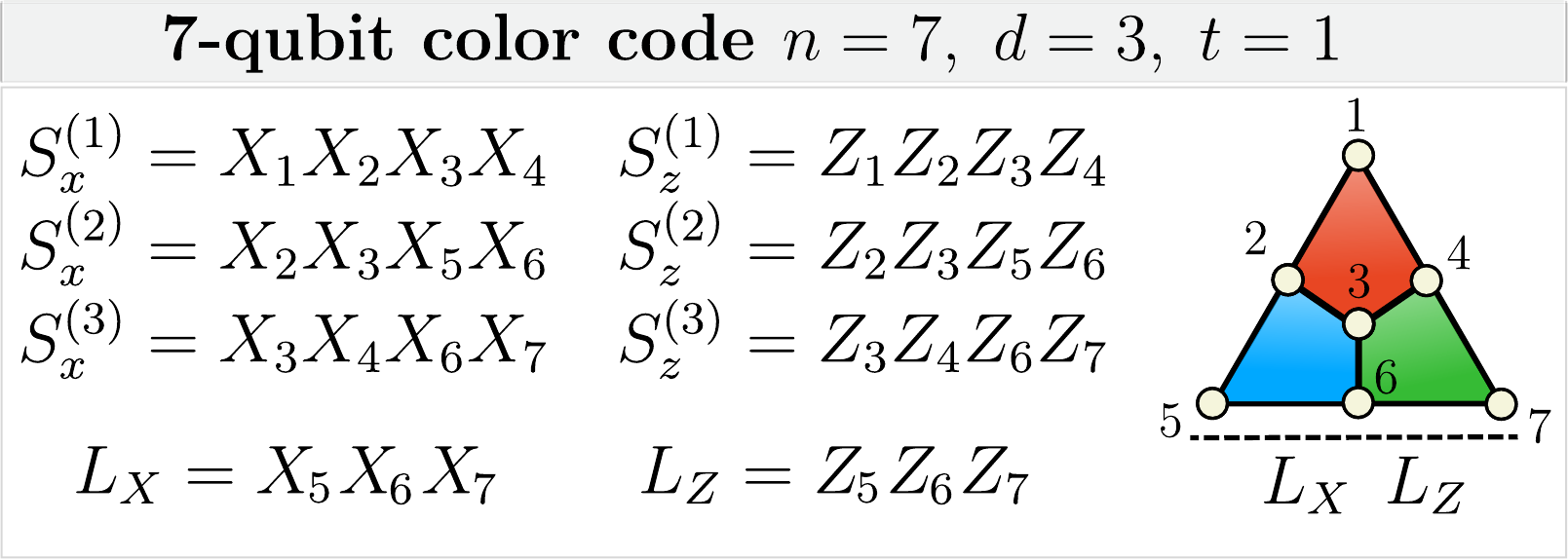}
\\
\caption{\textbf{The  7-qubit distance $d=3$ colour code:} 
We encode one logical qubit in $n=7$ physical qubits.  The qubits are represented by the dots in the vertices, the logical operators $L_X$ and $L_Z$ act on the edge of the triangle, and the stabiliser checks $S^{(i)}$ correspond to the plaquettes. The code can correct up to $t=1$ arbitrary fault on any of the seven physical data qubits.     }
\label{fig:colorcode}
\end{centering}
\end{figure}

Let us now briefly summarise a few central properties of the 7-qubit colour code~\cite{bombin-prl-97-180501, steane-prl-77-793}, which are relevant for the QEC protocol studied below. With this code (see Fig.~\ref{fig:colorcode}) it is possible to store and manipulate $k=1$ logical qubit redundantly encoded in entangled states distributed over $n=7$ physical qubits. The code has a logical distance of $d=3$ and, therefore, one can detect and correct at least $t=\left \lfloor{\frac{d-1}{2}}\right \rfloor =1$ arbitrary fault (phase and/or bit flip fault) on any of the seven physical data qubits. The code belongs to the family of CSS codes \cite{calderbank-pra-54-1098,steane-prl-77-793}, allowing one to independently detect and correct bit- and phase-flip faults. These faults are identified by measuring a set of three weight-four $X$-type and three $Z$-type parity checks. The measurements indicate parity violations resulting from the bit and phase flips and are used to deduce the presence of these faults.  

Logical states are encoded in the code space, which is defined as the simultaneous eigenspace of eigenvalue +1 of all six stabiliser generators listed in Fig.~\ref{fig:colorcode}. Logical qubits employing larger  distance codes, and thus allowing for the correction of multiple errors, can be constructed by encoding a logical qubit in larger 2D lattice structures involving more physical qubits \cite{bombin-prl-97-180501}. 

One of the main features that distinguishes the colour code from other codes, like, e.g. Kitaev's surface code \cite{dennis-j-mat-phys-43-4452}, is that the colour code permits the transversal, i.e.~bit-wise realisation of the entire Clifford group \cite{bombin-prl-97-180501, nielsen-book}. This property facilitates a fault-tolerant application of the logical gate operations, as applying the logical gate through the application of single-qubit unitaries prevents an uncontrolled propagation of errors through the circuit. To complete a universal set of logical gate operations, one needs to complement the Clifford operations with a single non-Clifford gate.  A common option is the $T$-gate \cite{nielsen-book},    which can be implemented in the 7-qubit colour code by magic-state injection \cite{s-bravyi-pra-71-022316} via a teleportation process between the register of system qubits and an ancilla qubit, through the preparation of a $T$-state and the growing of a logical qubit \cite{lattice_surgey_cnot_color_code}, or by code-switching techniques \cite{1509.03239,1509.04255,1512.04193}.

\begin{figure}[!ht]
 \begin{centering}
  \includegraphics[width=0.95\columnwidth]{ 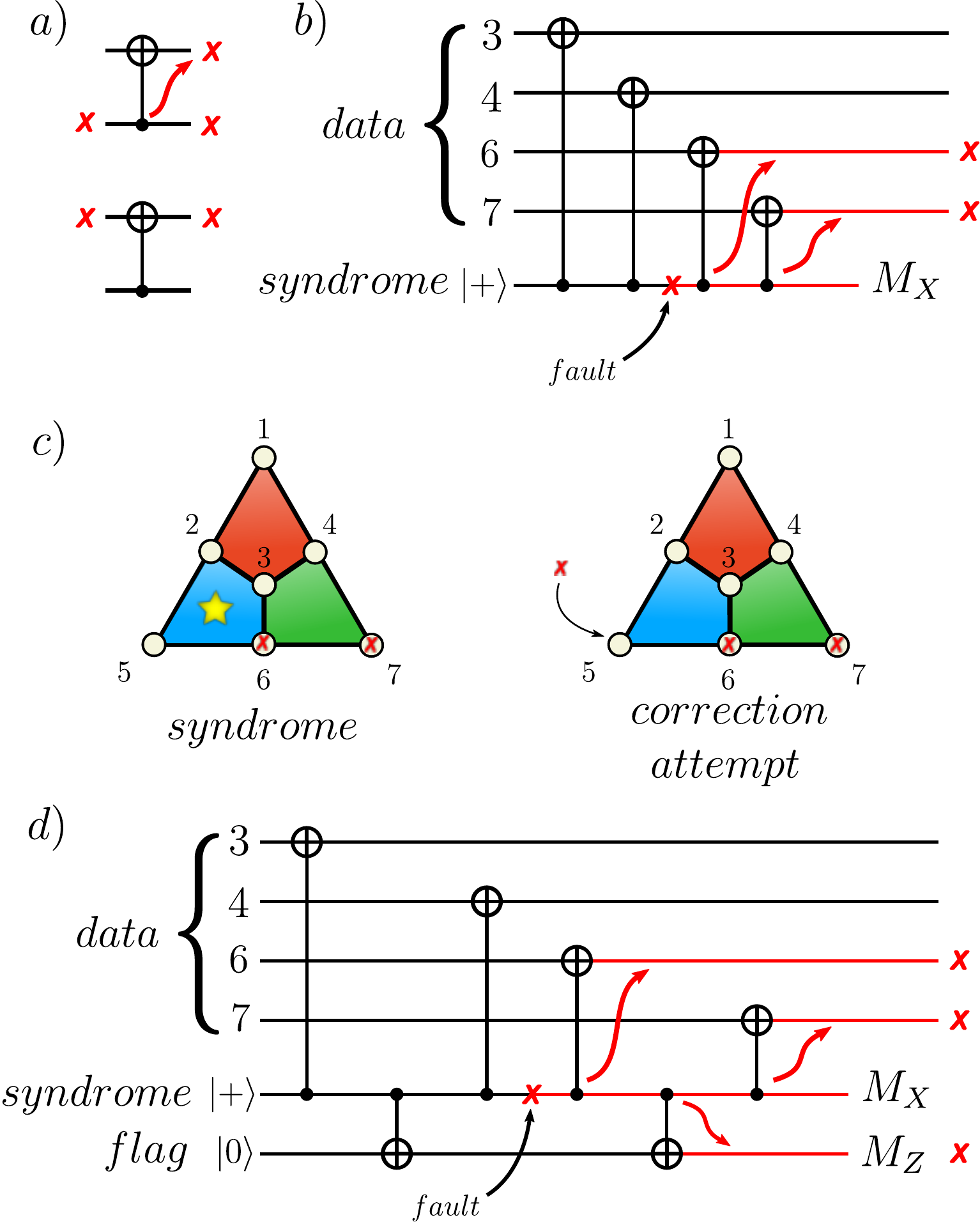}\\
  \caption{\textbf{Error propagation in the syndrome readout circuit.}\textbf{ a)} Bit flip error propagation through a CNOT gate: an X error in the control qubit propagates to the target qubit, but it does not propagate from the target  qubit to the control qubit. \textbf{b)} Standard stabiliser measurement of $S_x^{(3)}$ (green), where a single ancilla is used to measure the stabiliser. A single fault on the ancilla can propagate through the CNOTs and result in two X faults on the data qubits (6 and 7). Such faults that grow due to circuitry and trigger logical errors are called hook errors.  \textbf{c)} Syndrome readout: (left) after measuring all six stabilisers, only the blue stabiliser (marked with a star) will be excited by the two faults. (right) The correction corresponding to that syndrome would be to apply an $X$ operation to qubit 5, which would complete the logical $L_X=X_5X_6X_7 $ operator and lead to a logical-$X$ error. \textbf{d)} Stabiliser readout with a flag qubit: hook errors can be detected by using an additional ancilla (flag qubit) in the stabiliser readout. If a flag qubit signals a fault, another round of syndrome readout is run, and the information obtained by the flag and the syndrome information can be used to identify and correct the error \cite{singleflag-pra-brown}.}
  \label{Fig:nonFTparity}
\end{centering}
\end{figure}

A challenge of colour codes is that they are not naively fault-tolerant when using standard parity check circuits in which a single ancilla is used for a stabiliser measurement. Whereas fault-tolerant stabiliser readout with single ancilla qubits is possible in surface codes, other codes, including colour codes, require more sophisticated ancilla constructions. That is, for standard parity check circuits, some weight-one faults spread to weight-two faults that lead to a logical error for the 7-qubit colour code, as can be seen in the example in Fig.~\ref{Fig:nonFTparity}.
To overcome this problem, various schemes have been developed to make the stabiliser readout fault tolerant~\cite{shor_ft_qec,aliferis_ft_qec,steane_ft,knill_ft}. However, these schemes require a substantial resource overhead, as they are based on the use of multiple qubit ancilla states which need to be prepared and verified, typically in multi-partite entangled states~\cite{shor_ft_qec,aliferis_ft_qec}.

In more recent work, it was shown that an FT stabiliser measurement could be achieved with only one additional ancilla qubit, the so-called  flag qubit (Fig.~\ref{Fig:nonFTparity}). In a nutshell, the idea is that one ancilla is used for the stabiliser measurement while a second ancilla acts as a flag to indicate the possible occurrence of dangerous correlated faults that can lead to logical errors. In one QEC cycle, the stabilisers are measured sequentially, until one of the ancilla measurement signals a fault. When this occurs, the six stabilisers are measured again. Using the information of these measurements and the flag qubit, it is possible to infer and correct the  error, preserving  the FT character. 

In a recent work \cite{Reichardt-new}, a new fault-tolerant scheme is presented that allows for the parallel flagged measurement of stabilisers (Fig. \ref{fig.qecc.f2}). In this scheme, each ancilla is used to measure one of the stabilisers, but at the same time acts as a flag to indicate potentially dangerous faults happening on the other ancillas, which otherwise could propagate onto several data qubits and cause a logical error. If a fault is detected, all the stabilisers are measured again to obtain the necessary information to identify the fault. This method allows the simultaneous measurement of three stabilisers in parallel. The new flag-based scheme has the advantage of reducing the number of measurement rounds since it allows parity-check measurements to be performed in parallel. This reduces the time to perform a QEC cycle and the time during which idle qubits undergo decoherence, which is particularly advantageous in platforms with short coherence times and slow measurements.

\begin{figure*}
\begin{centering}
\includegraphics[width=1.82\columnwidth]{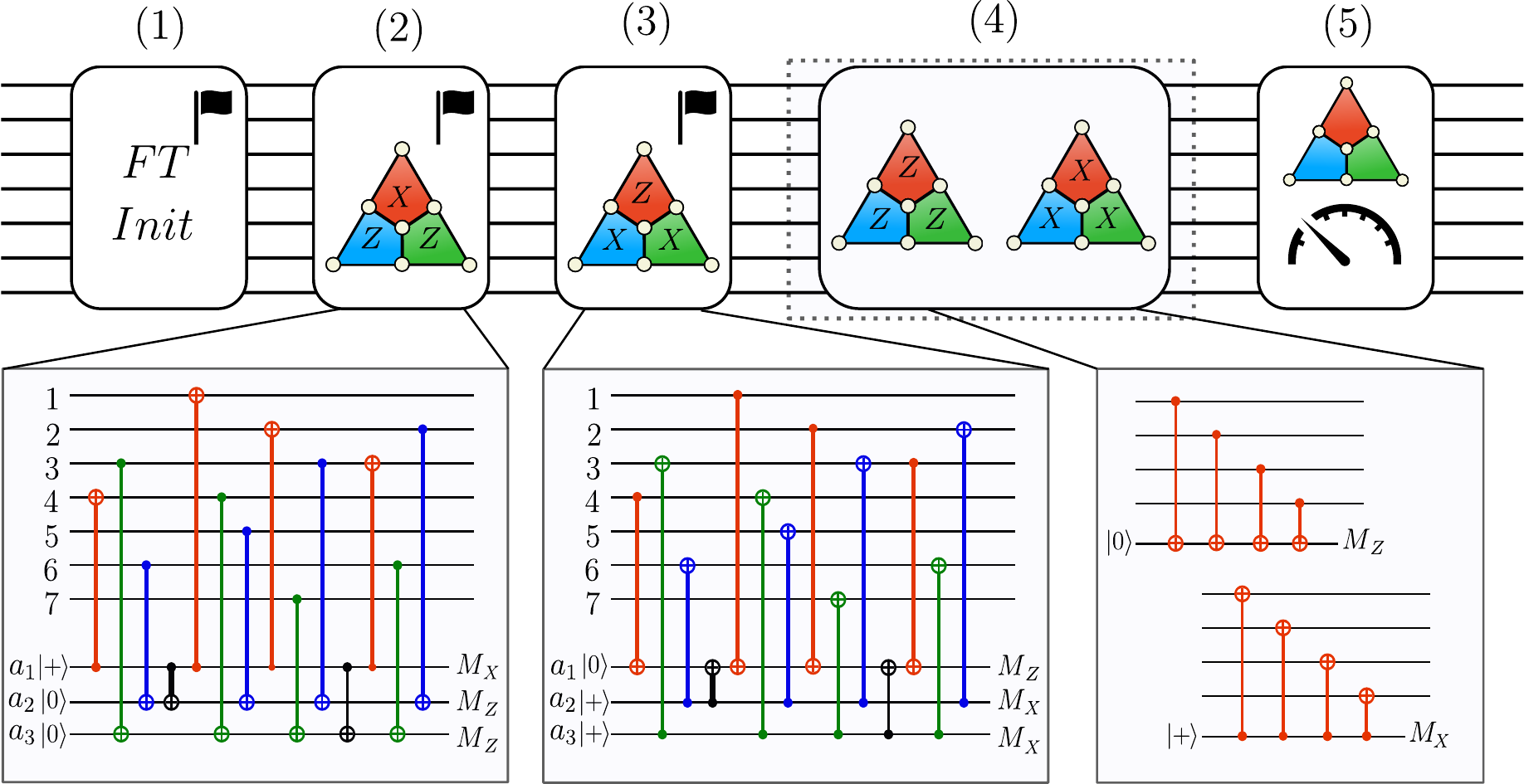}
\caption{\textbf{ QEC protocol:} the QEC round used to estimate the logical error probability works as follows: first, we initialize the state using a FT protocol \textbf{(1)} using a single ancilla as a flag \cite{1810.09199}. If the flag is raised, the state is discarded, and we restart the protocol. Then, we begin the syndrome extraction protocol:  \textbf{(2)}  parallel flagged syndrome extraction of the stabilizers $S_x^{(1)}$, $S_z^{(2)}$ and $S_z^{(3)}$ (circuit redrawn from Ref. \cite{Reichardt-new}). If no flag is raised, we proceed to \textbf{(3)} and measure the stabilizers $S_z^{(1)}$, $S_x^{(2)}$ and $S_x^{(3)}$ using parallel flagged readout. If any flag was raised during \textbf{(2)} or \textbf{(3)}, we realize a second full round of stabilizer readout without flags. Using three ancillas, we can measure simultaneously groups of three stabilizers. Finally, \textbf{(5)} the state is measured. We can apply the recovery operation corresponding to the syndrome and check if the protocol ended in success or failure. In the figure, the circuits are shown using CNOT gates. For the simulations, we compiled the circuits into MS gates to adapt them for the trapped-ion universal gate set. An example is shown in Fig. \ref{fig.qecc.f2} for the circuit in \textbf{(2)}.
}\label{fig.qecprotocol}
\end{centering}
\end{figure*}

\begin{figure}
\begin{centering}
\includegraphics[width=\columnwidth]{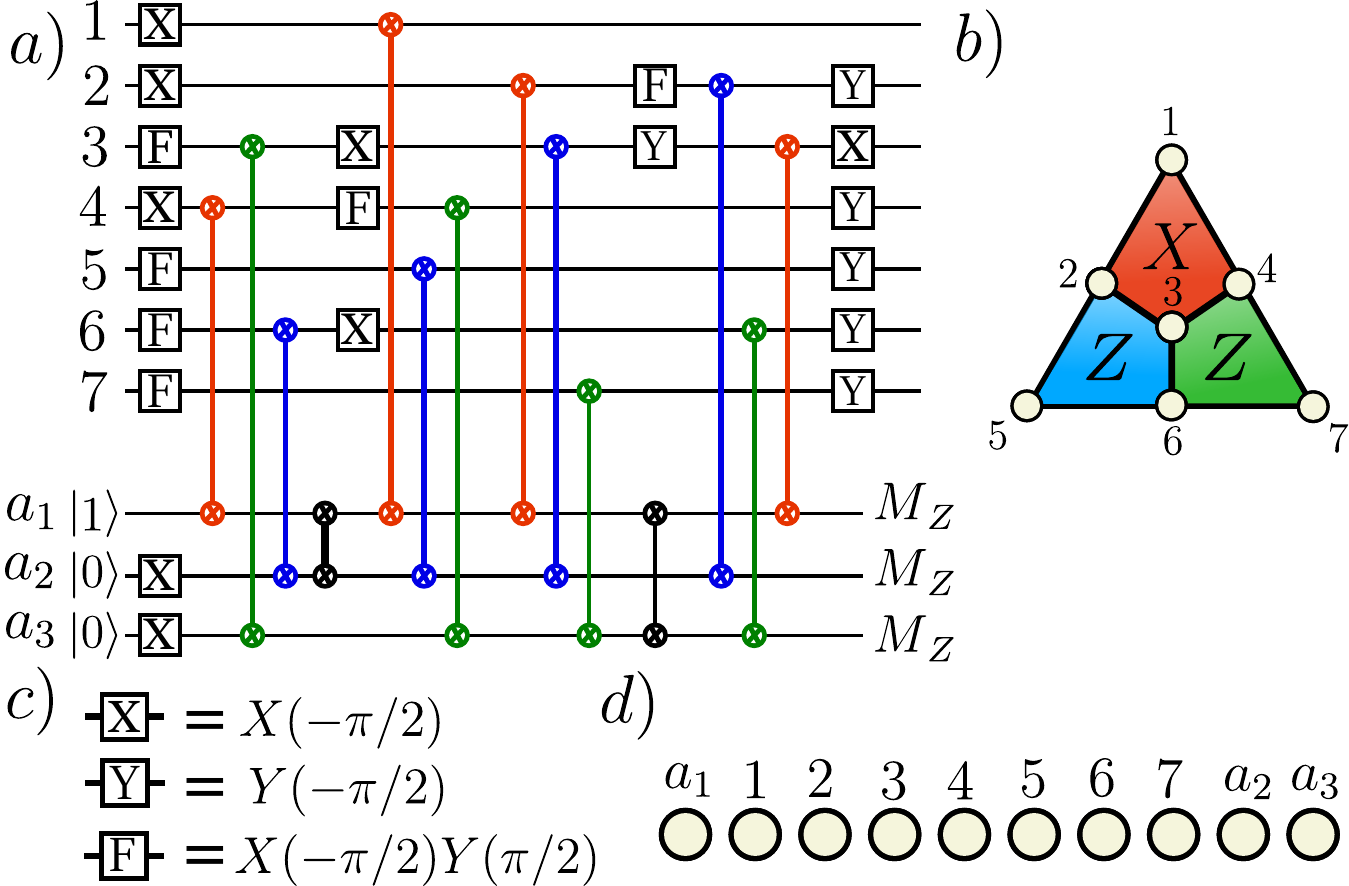}
\\
\caption{\textbf{ Circuits for parallel syndrome extraction:}
 \textbf{ a)} Circuit for the parallel and FT measurement of the stabilisers $S_x^{(1)}$, $S_z^{(2)}$ and $S_z^{(3)}$ via ancillas $a_1$, $a_2$ and $a_3$, respectively, as shown in \textbf{b)}. The circuit, shown in Fig. \ref{fig.qecprotocol} with CNOTs, has been compiled into MS gates with the addition of single-qubit gates, shown in \textbf{c)}. A similar circuit can be used to measure fault-tolerantly the stabilizers $S_z^{(1)}$, $S_x^{(2)}$ and $S_x^{(3)}$ by compiling the circuit with CNOTs shown in Fig. \ref{fig.qecprotocol}. Using those two circuits, it is possible to measure all 6 stabilizers with 3 ancillas in only two sets of operations.   \textbf{e)} Sketch of the order of the qubits in the ion string. 
}
\label{fig.qecc.f2}
\end{centering}
\end{figure}

Finally, it is important to note that, from a decoding standpoint, the new QEC protocols do not impose any significant increase in decoder complexity for the operation of a 7-qubit colour code logical qubit. Essentially, depending on whether and which ancilla qubits give non-trivial measurements, e.g.~a small set of lookup tables can be used to identify the suitable correction \cite{Reichardt-new}.

\section{\bf The adversity of bare crosstalk on QEC\label{sec.problem}}

In this section, we analyse the impact of crosstalk on the performance of the 7-qubit colour code using numerical simulations. We simulate individual rounds that begin with a fault-tolerant state preparation using a single ancilla \cite{1810.09199}, followed by a round of flagged QEC in which we measure the stabilisers using the flag scheme previously described in Sec.~\ref{sec.background}. If an error is detected, we do a second measurement of the stabilisers, without flags. We then apply a correction based on the information from the stabiliser measurements using a lookup table. Finally, we measure all the qubits to check if a logical error has happened, applying a classical round of error correction if needed. We simulate multiple iterations of this experiment to estimate the logical error rate of the circuit for each set of error parameters (see Fig. \ref{fig.qecprotocol}). 

\begin{figure}[!ht]
	\begin{centering}
        \includegraphics[width=8.5cm]{ 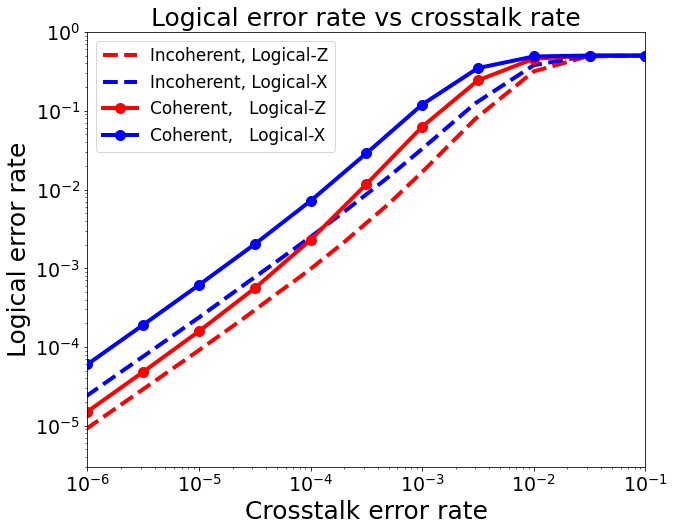}
		\\
		\caption{ \textbf{Logical error rates for coherent and incoherent crosstalk errors.}  Here, we estimate the logical error rate using a limited error model that includes crosstalk errors only and set all other error sources to zero. We measure the logical-$X$ (logical-$Z$) error rate by simulating a QEC round on the state $\ket{+}_L$ ($\ket{0}_L$) (colours blue and red in the figure, respectively).
		We  compare the results of simulating the crosstalk errors coherently (using state vector simulations and the crosstalk model on Eq. \ref{eqCrosstalk}, solid lines) or incoherently (using Pauli propagation simulations and the crosstalk error model on Eq. \ref{eq:Kraus_incoherent}, dashed lines). The results show that the logical error rates for coherent simulations are worse than the incoherent version by a factor of about 3, but that this difference remains approximately constant as  crosstalk is increased.
		The error bars for the points are smaller than the size of the marker.} 
		\label{Fig:singleflag_cvsi_onlyc}		
	\end{centering}
\end{figure}

First, we study the relevance of the coherent versus incoherent modelling of crosstalk errors. To avoid unnecessary complications, we note that this question can already be addressed by limiting the error model to crosstalk errors only. We  run state-vector numerical simulations to explore the QEC protocol's performance under coherent errors in Eqs.~\eqref{eqCrosstalk} and~\eqref{eq:crosstalk-starkshift-coherent} and compare the results with those stemming from stabiliser simulations for the incoherent error model of crosstalk in Eqs.~\eqref{eq:Kraus_incoherent} and~\eqref{eq:Kraus_incoherent_ac}. The results shown in Fig. \ref{Fig:singleflag_cvsi_onlyc} show that the logical error rates for the coherent model of crosstalk errors are on the same order of magnitude but, generally, slightly larger than those obtained using the incoherent crosstalk model. A linear fit in the low crosstalk regime, namely for $p_c<10^{-4}$, shows that the scaling of the logical error    rate with the crosstalk error rate is linear, as expected, and that the coherent version of crosstalk is a factor of $3.0\pm0.3$ larger than its incoherent counterpart. This indicates that the incoherent approximation of the crosstalk errors underestimates expected logical error rates, overestimating the QEC capabilities and leading to better pseudo-thresholds when the full error model is considered, as discussed below. Once this comparative analysis has been performed, let us analyse the performance of the 7-qubit colour code, using the flag scheme described  previously, under the full error model that incorporates all other microscopic error sources in addition to the  crosstalk noise. We run simulations for different values of the crosstalk error rate to identify a critical value of crosstalk errors below which it is possible to reach the break-even point of ``quantum logical advantage'', namely when the logical error rate is lower than the leading physical error rate in the typical circuits. As discussed previously for trapped-ion implementations, the largest physical error rate is the entangling MS gates.  We only consider one ion string ordering, in which ions in positions 2-8 correspond to qubits 1-7 of the Steane code, and ions 1, 9 and 10 correspond to the ancilla qubits (see Fig. \ref{fig.qecc.f2}).   Here, we focus on the analysis of the performance of the logical state $\ket{{+}}_{\rm L}=(\ket{{0}}_{\rm L}+\ket{{1}}_{\rm L})/\sqrt{2}$, as it is typically this state which performs worse under the effects of noise and imperfections. Thus, our simulations are conservative for the eventual goal of sustaining long-lived logical qubits.

 We use the open-source software known as PECOS \cite{1812.04735,PECOSGit} to perform the numerical simulations in this work. PECOS is a Python framework for studying, developing, and evaluating QEC protocols. Concepts such as QEC protocols, error models, and simulators are decoupled from each other. This allows us to specify our protocol circuitry and error models separately and choose the most appropriate simulator. For incoherent numerical studies, we utilised the stabiliser simulator included with the software; however, for coherent simulations, we ran state-vector simulations using ProjectQ \cite{1612.08091,1604.01401} as a backend to PECOS. While the coherent simulations provide a more precise estimate of a quantum system's behaviour, we want to emphasise the importance of the stabiliser simulations. These incoherent simulations provide fast results that allow for a quick evaluation of the behaviour of a given process. For many practical applications, this allows faster informed decisions and a better workflow.

\begin{figure}[!ht]
\begin{centering}
\includegraphics[width=0.95\columnwidth]{ 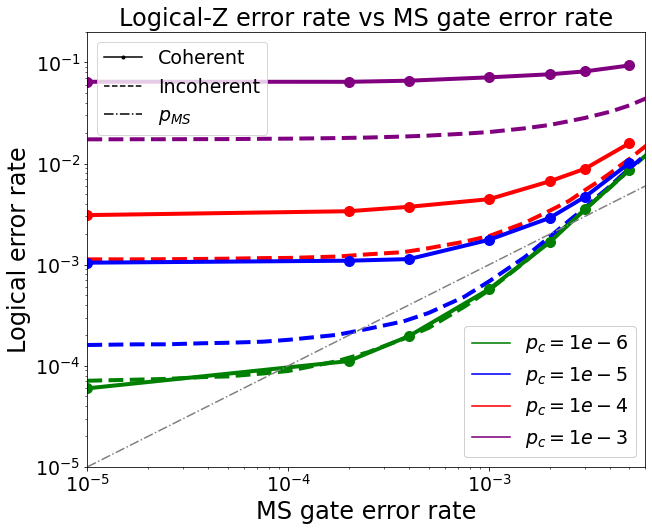}
\caption{
\textbf{Pseudo-threshold as a function of the entangling-gate error rate  for specific  crosstalk errors:} 
 We show the results, on a log-log scale, for preparing and sustaining logical $\ket{+}_{\rm L}$ state.    Note that for low MS-gate errors, the curves flatten. This is because, for each line, only the MS-gate errors vary and other error rates such as crosstalk, idle, single-qubit gate, initialisation, and measurement errors are held fixed. From the results of this plot, we see that if the crosstalk error-rate is at     $10^{-6}$ or better, then a pseudo-threshold exists. Furthermore, a pseudo-threshold of approximately $2.7 \times 10^{-3}$ can be obtained.   In addition to the MS gate errors, the simulation includes other sources of errors with the following rates: single-qubit errors ($p_{s}=10^{-5}$), state preparation and measurement errors ($p_{sp}=p_m=10^{-4}$), uncorrelated dephasing with coherence time $T_2 = 2.2s$, amplitude damping and leakage, with a lifetime $T_1 = 1.1s$ of the metastable qubit state $\ket{1}$. These parameters are used in the remaining simulations in the paper. The details on these error models are described in Section \ref{sec.noise}. The floor levels of the logical error rate observed for low MS-gate error rates stem from the damaging effects of these additional error sources.  The error bars for the points are smaller than the size of the marker.  }\label{fig:logZcoherent}
\label{Fig:ThreeChecksNoIgor}
\end{centering}
\end{figure}

The simulation results shown in Figure \ref{Fig:ThreeChecksNoIgor} represent the performance of QEC under a comprehensive error model that aims to include the relevant noise sources that affect near term, static-string ion trap devices, as described in Section \ref{sec.noise}. While the main control parameters are the crosstalk and MS-gate error rates, the simulation includes other sources of errors with the following rates: single-qubit errors ($p_{s}=10^{-5}$), state preparation and measurement errors ($p_{sp}=p_m=10^{-4}$), uncorrelated dephasing with coherence time $T_2 = 2.2s$, amplitude damping and leakage, with a lifetime $T_1 = 1.1s$ of the metastable qubit state $\ket{1}$. The floor levels observed on the logical error rate for low values of the MS-gate error rate depend on the fixed values of these additional error sources. These values of the parameters are used in the remaining simulations in the paper.

As can be seen in  Fig.~\ref{Fig:ThreeChecksNoIgor}, we find that for a crosstalk error-rate of    $p_c=10^{-6}$, or below, a pseudo-threshold can be obtained     when the entangling MS-gate error lies below $p_{\rm MS}\lesssim 2\cdot 10^{-3}$, a value that is at reach for current technology. Let us also remark that, contrary to idealised noise models with a single error parameter, the regime of quantum logic advantage is contained within a range of physical entangling-gate error rates. If the MS-gate error is reduced well below the levels of the other noise sources, it will no longer be possible to demonstrate quantum logic advantage by comparing the logical error rate with the MS-gate error rate, as the other sources of noise will be the leading ones. This leads to a plateau-like levelling out of the logical error rate, similar to the behaviour observed in Ref. \cite{Brown:2019}. This clarifies the inherent complexity of having a single well-defined threshold for realistic multi-parameter noise models, which justifies the search for other metrics to quantify the benefit of QEC~\cite{Bermudez2017,1810.09199}.   Although the required MS-gate error level is realistic for current state-of-the-art architectures, reducing crosstalk error rates to values on the order of $10^{-5}$-$10^{-6}$ might be challenging using current optical addressing techniques. This difficulty motivates using  refocussing techniques, as an additional circuit level technique to further suppress residual crosstalk. As we will show in the following sections, the application of a  refocussing scheme will allow for a reduction of crosstalk levels by several orders of magnitude, leading the scheme back to a beneficial QEC regime.  Further reduction of crosstalk to values below $c\simeq 10^{-7}$ does not improve the results, as its effects become negligible with respect to the other error sources present in the simulations. Although crosstalk errors still have the capability to break FT in the Steane code, as a single crosstalk fault event can lead to a non-correctable error, it stops being the limiting factor for the performance of QEC in the experimentally relevant regime.

\section{\bf Active suppression of crosstalk for QEC\label{sec.results}}

\subsection{ Refocussing and residual noise\label{sec. refocussing}}
\begin{figure}[!ht]
\begin{centering}
\includegraphics[width=0.99\columnwidth]{   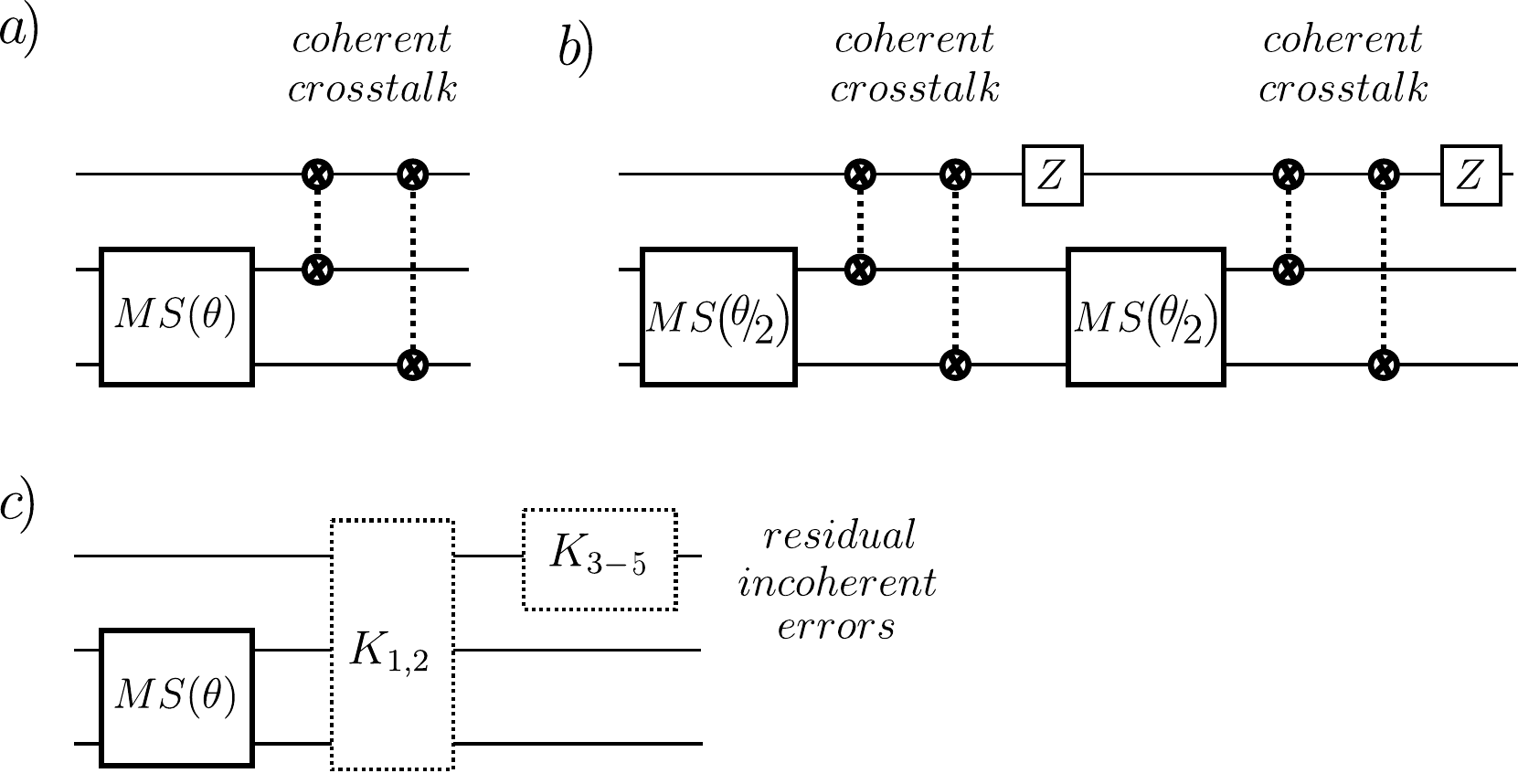}
\\
\caption{\textbf{Sketch of the crosstalk   refocussing scheme:}, illustrated for the compensation of crosstalk with a single neighbouring ion.\textbf{ a)} The effect of the crosstalk are coherent $X_GX_N$ partial entangling operations between each of the gate ions and the spectator ion, shown in dashed lines. \textbf{b)} The   refocussing sequence decomposes the total operation of the MS gate into two steps, with each MS gate corresponding to an MS gate of half the entangling angle  of the target MS gate. Between the two steps, We apply a single-qubit $Z$ rotation ($\pi$-pulse) to the spectator ion. This will lead to the cancellation of the coherent crosstalk built up between the gate qubits and the spectator qubit over the entire sequence, whereas on the two gate qubits the intended complete entangling operation (MS) is realised. \textbf{c)} As discussed in the main text, residual errors in the gates employed in the   refocussing sequence are expected to lead to residual, strongly suppressed and now incoherent residual crosstalk interactions of $X_GY_N$ type, as well as weak single-qubit dephasing on the spectator qubit(s).  Also, use notation $XX_ij(\theta/2)$ to be consistent with the introduced in the text.}
\label{Fig:MSgate  refocussing}

\end{centering}
\end{figure}

As we have seen in the previous section, crosstalk errors can have a damaging effect on a logical qubit's QEC performance. Under the first crosstalk error model discussed above, the leading order effect is a coherent partial entangling operation between all combinations of gate qubits and neighbouring qubits, as described in equation \ref{eqCrosstalk}. Due to this error's coherent nature, one can use composite pulse schemes to refocus these undesired interactions caused by crosstalk, like the scheme shown in Fig.~\ref{Fig:MSgate  refocussing}. The scheme we propose  for the compensation of the coherent crosstalk error is a simple  refocussing sequence, 
similar to the pulse sequences used in \cite{Leung_2000,Hayes_2014} to remove unwanted couplings and generate arbitrary Ising-like interactions. Our refocussing technique 
 consists of the application of {\it (i)} a half-entangling gate (MS/2)     between the pair of active ions $(i,j)\in G$, which corresponds to $XX_{ij}(\theta/2)$ in Eq.~\eqref{eq:MS_gate} for $\theta=\pi/2$, {\it (ii)} a $\pi$-rotation about the $Z$ axis on (all) the spectator ions(s) $n\in  neig\{G\}$, which corresponds to $Z_n(\theta)$ in Eq.~\eqref{eq:single_qubit_roattion} with $\theta=\pi$. {\it (iii)} A second half-entangling gate (MS/2)    on the active ions, and {\it (iv)}  final $\pi$-rotation(s) about the $Z$ axis of the spectator ion(s). 
It is straightforward to verify that, if implemented flawlessly, this composite sequence cancels the crosstalk interaction between gate and spectator qubits, while it realises the intended MS gate on the gate qubits.

To obtain an estimate of the remaining, uncompensated errors due to crosstalk in the  refocussing scheme, we consider a simple noise model. In this model, errors in the single-qubit  refocussing pulses suffer stochastic amplitude or phase fluctuations, resulting in  standard depolarising noise. The dynamics of the imperfections in the $\pi$-pulses used for the  refocussing can be written as 
 
 \begin{equation}
     \dot{\rho} \simeq  \sum_{i = 1}^{3}  \tfrac{1}{3}\expval{\epsilon_Z^2}  \left(\sigma^{(i)}_n \rho {\sigma^{(i)}_n}^\dagger - \frac{1}{2} \left\{ {\sigma^{(i)}}^\dagger \sigma^{(i)}_n , \rho \right\}\right)
 \end{equation}
 
where $  \epsilon_Z$ is a stochastic variable with probability distribution  $p( \epsilon_Z)$  and $\sigma^{(i)}$ are the Pauli matrices, and $\rho$ denotes the density operator of the system. This can be associated with a single-qubit error rate of $p_s= \expval{\epsilon_Z}^2 = \sigma_Z^2$, assuming that the distribution of the errors in the single-qubit $Z$ gates follow a Gaussian distribution with standard deviation $\sigma_Z$, and there are no systematic errors
 
 \begin{equation}
 \int d \epsilon_Z p( \epsilon_Z) \epsilon_Z = 0.
 \end{equation}
 
We can then obtain the following Kraus    decomposition~\eqref{kraus_decomp} for the noise channel by integrating the dynamics of the system, and averaging over the stochastic noise. In particular, we find that the Kraus operators will depend on the standard deviation that characterises the residual crosstalk error. 

One can show that, to leading order, the dynamics of the system of    active and spectator qubits can be described by incoherent dynamics. The leading order Kraus operators for the residual errors of the entire  refocussing sequence are

\begin{align}
K_0 = &\sqrt{1-2p_s}  \id_n,      & K_3 = &\sqrt{2\frac{p_s}{3}}  X_n,\\ \nonumber
K_1 = &\sqrt{\frac{p_s}{3}}  Z_n U_{CT}, & K_4 = &\sqrt{\frac{p_s}{3}}  Z_n, \\             \nonumber
K_2 = &\sqrt{\frac{p_s}{3}}  Y_n U_{CT}  & K_5 = &\sqrt{\frac{p_s}{3}}  Y_n, \nonumber
\end{align}
where $U_{CT} = \exp(-i \theta \epsilon_{CT} X_nX_g)$ is the unitary from the original crosstalk error between the active qubit $g$ and the neighbour qubit $n$. Here, the operators $K_3$ to $K_6$ correspond to standard depolarising error channel. The operators $K_1$ and $K_2$ correspond to two-body jump operators describing residual \textit{incoherent} crosstalk between gate and spectator ions. Note, however, that crucially the rate at which these residual errors occur, $p_{CT}^{residual}= p_s/3$, is strongly suppressed by the single-qubit error rate $p_s$. From this analysis, we can conclude that the application of the   refocussing scheme is expected to suppress the effects of the crosstalk noise by a factor of the single-qubit error rate, both for the coherent and incoherent cases. It therefore provides a powerful means to reach crosstalk error rates for which this error source starts to have a negligible influence on the QEC performance of the logical qubit, as our simulations discussed in the next section show.

\subsection{Performance of flag-based color-code QEC with active suppression of  crosstalk errors}
\begin{figure}[!ht]
\begin{centering}
\includegraphics[width=0.95\columnwidth]{  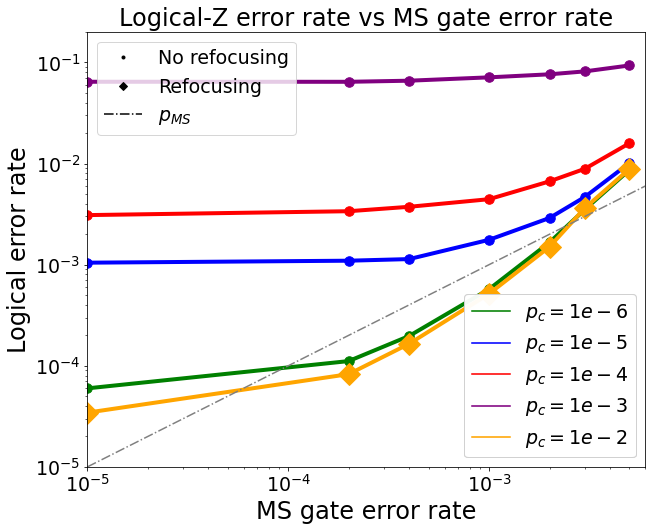}
\caption{\textbf{Pseudo-threshold for active suppression of  crosstalk: 
}  Here we determine the pseudo-threshold for the Steane code using parallel measurements of stabilisers and introducing the  refocussing scheme in the simulations. We overlap the results of the previous simulations, where we do not apply any crosstalk suppression technique. From the results on this plot, we see that at least four orders of magnitude reduction in the effect of crosstalk. The additional sources of noise modelled in the simulations use the same parameters as described in Fig. \ref{fig:logZcoherent}.  The error bars for the points are smaller than the size of the marker.} 
\label{Fig: refocussingResults}
\end{centering}
\end{figure}

We now turn to the study of the performance of the new FT flag-qubit based QEC protocol for parallel readout of three stabilisers \cite{Reichardt-new} with the  refocussing technique implemented for the MS gates. Here, as discussed in Subsection~\ref{sec.back.ions}, we use an error model that includes a Pauli channel approximation of the MS gate error model derived from the microscopic modelling, amplitude damping, leakage, errors in single-qubit initialisation, gates, and measurements as well as readout. In the simulations, the anticipated time durations are taken into account, with errors acting on idling qubits with a rate that scales with the corresponding duration. Additionally, and most importantly, for this study, we include coherent crosstalk and     apply the active suppression methods discussed above.

The simulations    displayed in Fig.~\ref{Fig: refocussingResults} show a clear window in which the QEC-corrected logical qubit, prepared in a logical $|+\rangle_L$ state, outperforms its physical counterpart and thus leads to quantum logical advantage. Importantly, we stress that this regime can be reached with a significant margin for MS gate fidelities of $10^{-3}$. Additionally, and as expected from the analysis and simulations in Sec.~\ref{sec.problem}, crosstalk error rates on the order of $10^{-3}$ have a negligible influence on the performance of the logical qubit given the level of suppression due to  refocussing, as suggested in  Sec. \ref{sec. refocussing}.

These simulations are consistent with the crosstalk suppression results found in the previous section. Together, these show a clear path towards realising a functional logical qubit in the regime of beneficial QEC for the static-string ion-trap approach.
\subsection{Performance of flag-based color-code QEC with the  Stark-shift crosstalk error model}

In this section, we study the effects of crosstalk using the second model  explained in Sec. \ref{sec.noise}, corresponding to the application of a power imbalance between the intensities of the red and blue sidebands to suppress the entangling interactions of crosstalk.   

\begin{figure}[!ht]
\begin{centering}
\includegraphics[width=0.95\columnwidth]{ 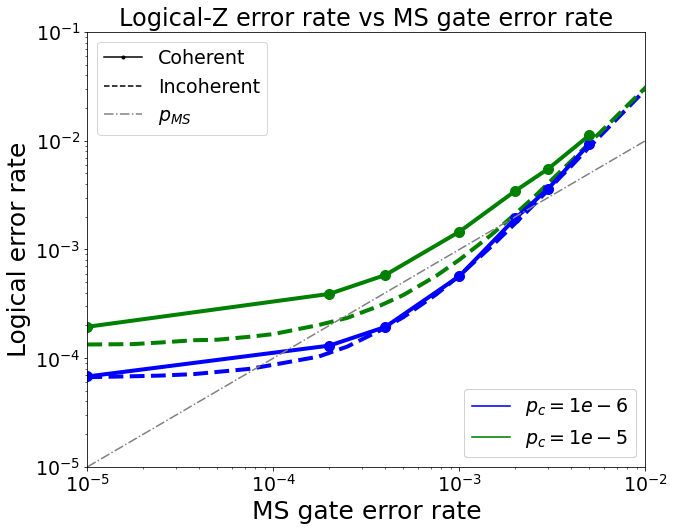}\\
\caption{\textbf{Pseudo-threshold for power-imbalanced suppression of  crosstalk:} A log-log graph comparison of the logical error-rates for different values of crosstalk, using the alternative crosstalk model with ac-Stark shifts on the neighbouring ions.   Here we determine the pseudo-threshold for the Steane code using parallel measurements of stabilisers as we vary MS-gate and crosstalk error-rates. We show the results for preparing a logical $\ket{+}$ state. The additional sources of noise modelled in the simulation using the same parameters as described in Fig. \ref{fig:logZcoherent}. The error bars for the points are smaller than the size of the marker.} 
\label{Fig:starkshift}
\end{centering}
\end{figure}

We simulate the resulting crosstalk errors both as coherent $Z$ rotations or as incoherent phase errors on neighbouring ions, with the results shown in Fig.~\ref{Fig:starkshift}.
From the simulations, we see that a pseudo-threshold can be found for crosstalk values on the order of $10^{-5}$ for coherent crosstalk, and on the order of $10^{-6}$ for incoherent crosstalk.

A problem for future work concerns the compensation of these new crosstalk rotations. As it is a systematic $Z$ rotation on the neighbours and does not create any entanglement, it could be possible to compensate it directly using single-qubit gates.
The question that remains to be answered is if the fidelity of the single-qubit gates is good enough for this type of compensation to be beneficial for QEC. It might be that these crosstalk levels are on the order of magnitude of the single-qubit errors, in which case the direct compensation of these errors might be worse than crosstalk itself.

\section{\bf Conclusion and outlook\label{sec.conclusion}}

In this work, we have presented a thorough analysis of flag-based colour-code QEC performance in ion traps using the static single-string approach. We have  used detailed and realistic error models that take into account various microscopic sources of noise, which we believe is crucial to provide a realistic estimate of the expected performance of near-term trapped-ion experiments. 

In previous works \cite{Bermudez2017,1810.09199},  a detailed study of the feasibility of small  topological trapped-ion  QEC codes under realistic models of noise  has been presented. These studies focused on mixed-species ion crystals in micro-fabricated traps, exploiting  shuttling and ion crystal reconfigurations for the implementation of QEC protocols. With the advent of improved laser focusing techniques for both single and two-qubit gates, it is nowadays possible to realize these small QEC codes  in a static string approach. Note, however, that the absence of shuttling requires a careful reconsideration of the error model: the close proximity of spectator ions in the string during gates addressed at active ions requires a careful consideration of  crosstalk noise, which was not considered  previously~\cite{Bermudez2017,1810.09199}. 
As shown by the results of our numerical simulations, bare crosstalk can compromise the demonstration of quantum logic advantage, as it places the regime of beneficial QEC within very small values of the crosstalk error rates, which are  out of the scope of current experimental technologies.
 As discussed in our work, the coherent nature of this noise allows us to design a  refocusing scheme to actively suppress crosstalk,    ultimately placing the pseudo-thresholds of quantum logic advantage within reach of near-term trapped-ion experiments. In this way, one can benefit from the simplicity of the static-string approach, and  achieve  quantum logic advantage  in analogy to the shuttling-based approach~\cite{Bermudez2017,1810.09199}.  

In conclusion, the overall message from this work on QEC with the  refocussing technique is very positive. As observed in the preceding sections, we have confirmed, both numerically and analytically,  that the  refocussing technique can dramatically suppress crosstalk noise. Therefore, placing us in a position of very strong confidence that we may meet the goal of realising a complete trapped-ion logical qubit in the beneficial QEC regime using the static single-string approach.

Following recent experimental results, we also introduce and study the impact of a second model for crosstalk noise on QEC performance, in which power imbalance in the MS gate generates Stark-shifts on the neighbouring ions. The results from this work indicate that this type of crosstalk has a smaller impact on QEC performance, and a pseudo-threshold can still be found for the anticipated experimental values even without applying any  refocussing sequence to compensate for these Stark-shifts.

We have focused our analysis on the 7-qubit colour code, using the parallel syndrome readout proposed in Ref.~\cite{Reichardt-new}.  Moving forward, it would be interesting to expand the study to other codes, namely, the surface code or higher distance colour codes. Another interesting study would be to combine the  refocussing scheme with other techniques to fight crosstalk, like the optimisation of the order of the ions to minimise the impact of the crosstalk events \cite{Brown:2019}. Further potential improvements can be found by operating a
two-species ion trap. This has the advantage that crosstalk between qubits encoded in different species is absent (beyond trivial systematic
light shifts that can be compensated), due to the different electronic level structure.

\section{Acknowledgements}


We acknowledge fruitful discussions with colleagues from the eQual and AQTION collaborations, and Davide Vodola. We thank the anonymous reviewer for useful comments. We gratefully acknowledge support by the EU Quantum Technology Flagship grant AQTION 820495. The research is based upon work supported by the Office of the Director of National Intelligence (ODNI), Intelligence Advanced Research Projects Activity (IARPA), via the U.S. Army Research Office Grant No. W911NF-16-1- 0070. The views and conclusions contained herein are those of the authors and should not be interpreted as necessarily representing the official policies or endorsements, either expressed or implied, of the ODNI, IARPA, or the U.S. Government. The U.S. Government is authorized to reproduce and distribute reprints for Governmental purposes notwithstanding any copyright annotation thereon. Any opinions, findings, and conclusions or recommendations expressed in this material are those of the author(s) and do not necessarily reflect the view of the U.S. Army Research Office. A.B. acknowledges support from the Ram\'on y Cajal program RYC-2016-20066, and CAM/FEDER Project S2018/TCS-4342 (QUITEMAD-CM) and the Plan Nacional Generación de Conocimiento PGC2018-095862-B-C22. We also acknowledge that computational resources were provided by Supercomputing Wales. 
\bibliographystyle{unsrtnat}
\bibliography{bib.bib}
\appendix

\section{Microscopic error  model for crosstalk\label{sec:appendix_A}}

This Appendix presents a detailed microscopic description of crosstalk in single- and two-qubit trapped-ion gates, which arise from residual illumination of spectator ions residing in the active ions' direct spatial vicinity for the corresponding gate. We now discuss the microscopic modelling of crosstalk, based on the underlying laser-ion coupling generating the gates, which provides us with a detailed understanding of the crosstalk error channels, both in their form and how their strength is related to the relevant experimental parameters.

\subsection{Coherent  crosstalk in single-qubit  gates\label{subsec:crosstalk_modeling}}

The effect of crosstalk in single-qubit gates can be modelled as follows: for resonant rotations ($X$ or $Y$-gates, or any rotation in the equatorial plane of the single-qubit Bloch sphere~\eqref{eq:single_qubit_roattion}), residual light on neighbouring ions causes undesired rotations of the spectator qubits in the same basis as the target active qubit

\begin{align}
& R_\mathrm{eff}(\Theta_\mathrm{eff}(t),  \phi_\mathrm{F}(t)) = \cos \frac{\Theta_\mathrm{eff}}{2} \id_2  \nonumber \\
& - \ii \, \sin \frac{\Theta_\mathrm{eff}}{2} \left( \cos \phi \hat{X}_\mathrm{eff}(t) + \sin \phi \hat{Y}_\mathrm{eff}(t) \right),
\end{align}

 where we have introduced an effective rotation angle determined by the residual attenuated intensity
 
\begin{equation}
\label{eq:ct_angle_1q}
\Theta_\mathrm{eff}(t) = \Theta \cdot \sqrt{\frac{I_{\mr r}(t)}{\langle I(t)\rangle}}.
\end{equation}

 This angle fluctuates around a fraction of the target value $\Theta$, which is selected by fixing the pulse duration and average intensity of the laser beams $\langle I(t)\rangle$ seen by the active qubit, due to the fluctuations of the residual intensity  $I_{\rm r}(t)$ seen by the spectator, which  can be reconstructed from the experimentally measured power spectral density~\cite{1810.09199}.   We note that laser phase fluctuations $\phi+\phi_\mathrm{F}(t)$ occur at a much slower timescale, and will lead to  gates  in a  rotated basis that slightly drifts from the target value $\phi$ as the quantum circuit proceeds, yielding
 
\begin{equation}
\begin{split}
\hat{X}_\mathrm{eff}(t) = \cos \phi_\mathrm{F}(t) X + \sin \phi_\mathrm{F}(t) Y  \\
\hat{Y}_\mathrm{eff}(t) = \sin \phi_\mathrm{F}(t) X + \cos \phi_\mathrm{F}(t) Y.
\end{split}
\end{equation}

As discussed in~\cite{1810.09199}, these fluctuations can be reconstructed from the power spectral density from a laser phase lock and Ramsey measurement. Note that crosstalk in single-qubit $Z$-rotations is quadratically smaller than that for $X$ or $Y$ rotations, since $Z$ rotations are implemented by AC Stark shifts, for which the effective rotation angle $\Theta_\mathrm{eff}$ is directly proportional to the residual intensity. 

Let us emphasise that there are well-established techniques to strongly suppress coherent crosstalk errors in single-qubit gates,  such as various dynamical decoupling sequences. One of the most simple composite pulse sequences to achieve ``addressing error correction" has been used routinely in ion traps, including, e.g., in the first colour code implementation experiments (see the supplementary information of Ref.~\cite{nigg-science-345-302} for a detailed discussion). \\

\subsection{  Crosstalk error channel for multi-qubit   gates\label{subsec:crosstalk_modeling_N_qubit_MS}}

 In this  part of the Appendix, we 
 introduce the effective \textit{entangling crosstalk error model} for $N$-qubit MS gates,  generalising~\cite{Bermudez2017}. We recall that  the procedure followed in~\cite{Bermudez2017} was to relate an analytical expression of the MS-gate  fidelity for the target GHZ-type state $\mathcal{F}=1-\epsilon_{\rm MS}$, including various sources of errors in the experiment, to the error rate $p_{\rm MS}$  of an effective multi-qubit quantum channel (e.g. depolarising channel). In this way, one gathers intuition about the contribution of physical error sources to the error rate of standard quantum channels  typically used in a QEC context, which can then serve to develop more sophisticated  models~\cite{1810.09199}.

Starting from~\cite{PhysRevA.62.022311}, we have built a microscopic model that accounts for crosstalk an MS gate and can incorporate  the crosstalk contribution $\epsilon_{\rm ct}$  to $\epsilon_{\rm MS}$, which will affect the spectator ions that will undergo dynamics resulting from residual couplings due the bichromatic  laser beams realising the MS gate. The various sources of error on the spectator ions can be grouped into two main types:

{\it (a) Coherent error contributions:} In an ideal situation, with no errors from dephasing, thermal phonon populations, or laser intensity fluctuations, an imperfect focusing of the MS-gate beams will affect the spectator ions due to two main sources. If the $N$-ion MS gate acts on the $\sigma^y$-basis with Rabi frequency $\Omega$, it will couple to $M$ spectator ions with a residual relative Rabi frequency $\{\Omega_j/\Omega\}_{j=1}^{M}$. The ideal bichromatic laser-ion interaction leads to a state-dependent force on the centre of mass (COM) mode

\begin{equation}
\label{eq:state_dep_force}
H_{\rm int}=-\frac{2\Omega\eta}{\sqrt{N+M}}\tilde{J}_ya_{\rm COM}\ee^{-\ii\omega_{\rm COM}t}\cos(\delta t)+{\rm H.c.},
\end{equation}

\noindent where we have introduced the  Lamb-Dicke parameter $\eta$, and the opposite detunings $\pm\delta$ of the bichromatic beams with respect to the blue- and red-sideband transitions, each of which  creates $a_{\rm COM}^\dagger$ or annihilates $a_{\rm COM}$  a vibrational excitation on the COM mode $\omega_{\rm COM}$ during the excitation of the internal state. Additionally, we have introduced the collective spin operator

\begin{equation}
\label{eq:coll_spin}
\tilde{J}_y=\frac{1}{2}\sum_{i=1}^N\sigma_i^y+\frac{1}{2}\sum_{j=1}^M\frac{\Omega_j}{\Omega}\sigma_j^y,
\end{equation}

\noindent which not only involves  the active ions $i\in\{1,\cdots N\}$, but also  the spectator ions $j\in\{1,\cdots M\}$ with residual light coupling. The ideal MS gate, after closing the phase-space trajectory, will induce a $\tilde{J}_y^2$ spin-spin interaction that also involves the spectator qubits. The dynamics will thus  depart from the target  fully-entangling MS-gate operation

\begin{align}
&\ket{0_1,\cdots,0_N}\otimes\ket{\psi_M^0}\to\ket{\Psi_{\rm id}}=\ket{{\rm GHZ}_N}\otimes\ket{\psi_M^0},
\end{align}
  which should prepare a GHZ-type state for the active ions $\ket{{\rm GHZ}_N}=\frac{1}{\sqrt{2}}\left(\ket{0_1,\cdots,0_N}-\ii(\ii)^N\ket{1_1,\cdots,1_N}\right)$, and leave the spectator qubits unaltered, e.g. in the initial state $\ket{\psi_M^0}=\ket{0_1,\cdots,0_M}$. Assuming that the crosstalk is small $\Omega_j\ll\Omega$, it is straightforward to evaluate how the MS gate will modify the spectator state $\ket{\psi_M^0}$, leading to $\mathcal{F}_{\rm ct}=1-\epsilon_{\rm ct}^{\rm MS}$, where 

\begin{equation}
\label{eq:coherent_rotation_ct_error_MS}
\epsilon_{\rm ct}^{\rm MS}=\frac{\pi^2}{4}N\sum_{j=1}^{M}\left(\frac{\Omega_j}{\Omega}\right)^{\!\!2}.
\end{equation}

Let us note that this crosstalk error occurs even in the absence of other error sources (i.e. for an ideal MS gate). We now introduce and discuss, one by one, these additional errors. The first one results from an off-resonant carrier excitation that appears in conjunction with the above state-dependent force of Eq.~\eqref{eq:state_dep_force}, namely

\begin{equation}
H_{\rm int}=2\Omega\tilde{J}_x\cos(\delta t)+{\rm H.c.},
\end{equation}

\noindent where  $\tilde{J}_x$ is a collective spin operator like Eq.~\eqref{eq:coll_spin}, but acting in the orthogonal $\sigma^x$-basis, and also containing the residual driving of the spectator ions. 
Generalising the perturbative calculations of~\cite{PhysRevA.62.022311}, one finds that the above crosstalk error gets an additional contribution

\begin{equation}
\epsilon_{\rm ct}^{\rm off}=\frac{1}{2}\left(\frac{\Omega}{\delta}\right)^2\sum_{j=1}^{M}\left(\frac{\Omega_j}{\Omega}\right)^{\!\!2}.
\end{equation}

Let us note that these crosstalk error contributions are coherent in nature, and can be undone using  refocussing techniques, as  discussed in Sec.~\ref{sec. refocussing}.  In the following, we will discuss crosstalk errors involving dephasing noise and thermal phonons, leading additionally to incoherent error sources that cannot be compensated by  refocussing.

{\it (b) Incoherent error contributions:}  The previous error sources assume that the laser-ion interaction only couples to the COM mode via the state-dependent linear term. However, in practice, there are additional terms that make the coupling deviate from this idealised limit and contribute to the error. For instance, there will be Debye-Waller factors that make the state-dependent force fluctuate depending on the thermal occupation of all the vibrational modes.  We then find that the infidelity due to these error source is

\begin{equation}
\epsilon_{\rm ct}^{\rm DW}=\frac{\pi N}{4}\sum_{n=1}^{N+M}\sum_{j=1}^M\left(\frac{\Omega_j}{\Omega}\frac{\eta^2_n}{N+M}\right)^{\!2}(2\bar{n}_n+1)^2,
\end{equation}

\noindent where we sum over all vibrational modes with thermal occupations $\bar{n}_n$ and Lamb-Dicke parameters $\eta_n$. 

In addition to the Debye-Waller factors, the phase-space trajectories may not close (especially for spectator modes), contributing to an incoherent error source due to residual spin-motion entanglement. This then leads to the following contribution to the MS-gate infidelity

\begin{equation}
\epsilon_{\rm ct}^{\rm loops}=\sum_{j=1}^M\sum_{n=2}^{N+M}\left(\frac{\Omega_j}{\omega_n}\right)^{\!2}\left(\eta_n\mathcal{M}_{j,n}\right)^2(2\bar{n}_n+1),
\end{equation}

\noindent where $\omega_n$ ($\mathcal{M}_{j,n}$) stand for the  eigen-frequencies (eigen-vectors) of the remaining collective vibrational modes.

Finally, there will be another incoherent contribution stemming from the dephasing and laser-intensity noise. The dephasing, to leading order, does not depend on the residual intensity of the MS laser beams and contributes with 

\begin{equation}
\label{eq:error_dephasing}
\epsilon_{\rm ct}^{\rm deph}=2N^2t_{\rm g}/T_2,
\end{equation}

\noindent where we have assumed that the phase noise acts collectively on the qubit register, and the coherence time is much longer than the gate time $T_2\gg t_{\rm g}$. For the laser-intensity fluctuations, characterised by a rate $\Gamma_I$, we find that the spectator ions with residual illumination suffer a crosstalk error contribution of

\begin{equation}
\epsilon_{\rm ct}^{\rm int}=\gamma_It_{\rm g}\sum_n\eta_n^2(\bar{n}_n+\half)\sum_j\left(\frac{\Omega_j}{\Omega}\right)^{\!\!2}.
\end{equation} 

Let us recall that one can use an approximate expression for all these sums, which partially simplifies the gate-infidelity expressions~\cite{PhysRevA.62.022311}. In any case, it is easy to realise that the total crosstalk contribution to the MS-gate  infidelity, obtained by adding all of the above terms,  can be expressed in terms of the infidelity $\epsilon_{\rm MS}$ of the MS-gate without crosstalk~\cite{Bermudez2017}  as follows

\begin{align}
\epsilon^{\rm ct}_{\rm MS}=&\sum_{j=1}^M\left(\frac{\Omega_j}{\Omega}\right)^{\!\!2}\left(\epsilon_{\rm MS}+\frac{\pi^2}{4}N +2N^2\frac{t_{\rm g}}{T_2}\right)\nonumber \\
&-2N^2\frac{t_{\rm g}}{T_2}.
\label{eq:error_total_MS_Nqubits}
\end{align}

Therefore, except for the dephasing contribution,  which is independent of the residual intensity, the crosstalk contribution to the infidelity is reduced by a small factor 

\begin{equation}
\label{eq:small_parameter}
\chi=\sum_{j=1}^M\left(\frac{\Omega_j}{\Omega}\right)^{\!2},
\end{equation}

\noindent due to the residual illumination of the spectator ions,  which will be   $\chi\ll1$ for sufficiently-good  addressing capabilities. This simple expression shows that the crosstalk  contribution of each source of error, from off-resonant carrier to thermal phonon populations, has the same relative weight as discussed at length for the MS gate without crosstalk~\cite{Bermudez2017}. All these contributions can then be used to extract the error rate $p^{\rm ct}_{\rm MS}$ that should be fed into a particular channel, such as a depolarising channel taking into account the crosstalk. For instance, if we restrict to a depolarising channel combining single-qubit and two-qubit operations~\cite{Bermudez2017}, considering now both active and spectator ions, the corresponding quantum channel is

 \begin{align}
\label{depolarising_channel_1_2_ct}
{\varepsilon_{\rm MS}}(\rho)&=(1-p_{\rm MS}-p_{\rm MS}^{\rm ct})\rho \nonumber \\
&+\frac{p_{\rm MS}}{N_{\rm MS}}\sum_{i=1}^{N}\sum_{\alpha=1,2,3}\sigma_{i}^\alpha\rho\sigma_{i}^\alpha \nonumber \\
&+\frac{p_{\rm MS}}{N_{\rm MS}}\sum_{i_1, i_2}\sum_{\alpha,\beta}\sigma_{i_1}^\alpha\sigma_{i_2}^\beta\hspace{0.2ex}\rho\hspace{0.2ex}\sigma_{i_1}^\alpha\sigma_{i_2}^\beta \nonumber \\
&+\frac{p_{\rm MS}^{\rm ct}}{N_{\rm ct}}\sum_{j=1}^{M}\sum_{\alpha=1,2,3}\sigma_{j}^\alpha\rho\sigma_{j}^\alpha \nonumber \\
&+\frac{p_{\rm MS}}{N_{\rm MS}}\sum_{j_1, j_2}\sum_{\alpha,\beta}\sigma_{j_1}^\alpha\sigma_{j_2}^\beta\hspace{0.2ex}\rho\hspace{0.2ex}\sigma_{j_1}^\alpha\sigma_{j_2}^\beta,
\end{align}

\noindent where $N_{\rm MS}=15N(N-1)/2+3N$ are all possible single- and two-qubit errors in the MS gate with $N$ active qubits, while $N_{\rm ct}=15(N+M)(N+M-1)/2+3(N+M)$ are all possible  single- and two-qubit errors in the MS gate with $N$ active qubits and $M$ spectator qubits. The final ingredient to relate the MS gate error rates for active $p_{\rm MS}$ and spectator $p_{\rm MS}^{\rm ct}$ qubits is to calculate the fidelity using the above quantum channel. One finds

\begin{align}
\mathcal{F}_{\rm MS}&=1-p_{\rm MS}\frac{N_{\rm MS}-N(N-1)/2}{N_{\rm MS}} \nonumber \\
& p_{\rm MS}=\epsilon_{\rm MS}\frac{N_{\rm MS}}{N_{\rm MS}-N(N-1)/2},
\end{align} 

\noindent which takes into account that, from all the possible $N_{\rm MS}$ single- and two-qubit errors, the   $\sigma_{i_1}^z\sigma_{i_2}^z$ errors do not affect the target GHZ state.  For the spectator ions, all the possible $N_{\rm ct}$ errors perturb the target state, such that 

\begin{equation}
\mathcal{F}_{\rm ct}=1-p^{\rm ct}_{\rm MS}\implies p^{\rm ct}_{\rm MS}=\epsilon^{\rm ct}_{\rm MS}.
\end{equation} 

As a summary,  we note that all of the additional  errors resulting from crosstalk in the MS gate, except for Eq.~\eqref{eq:coherent_rotation_ct_error_MS}, yield  a negligibly-small contribution. They are either due to the dephasing~\eqref{eq:error_dephasing}, which is often negligible for the short duration of the MS gate, or  they are proportional to the already-small error rate of the MS-gate  weighted by the additional small pre-factor~\eqref{eq:small_parameter}. Accordingly, the leading source of crosstalk errors will stem from the coherent contribution in Eq.~\eqref{eq:coherent_rotation_ct_error_MS}. The present error model is useful to make a quantitative comparison of the shuttling-based versus static-string performance of the QEC protocols discussed in Ref.~\cite{Bermudez2017}, assessing quantum logic advantage with both approaches by a microscopic description of the errors on the same footing.


\subsection{Sophisticated crosstalk noise for two-qubit gates}

As  the effective error channel~\eqref{depolarising_channel_1_2_ct} shows, entangling gates involving many qubits can break the fault-tolerance design principle, as an error in one of the data qubits  could potentially propagate to  other data qubits. To minimise this effect, we restrict the set of entangling gates in the circuits to two-qubit MS gates.  For this type of gates, the complete qubit and phonon degrees of freedom of the two-ion system can be efficiently simulated using numerical methods that include various sources of noise~\cite{1810.09199}. This has the advantage that, rather than relying on analytical calculations for the state infidelity fed onto phenomenological quantum channels, it becomes possible to extract the exact quantum channel by numerically averaging over  the stochastic noise processes. As shown in~\cite{1810.09199}, the main lesson is that the realistic channel for a noisy 2-qubit MS gate has more structure than the highly-symmetric depolarising channel. Unfortunately, generalising these exact numerical simulations to include crosstalk rapidly exceeds the computing capabilities, as each spectator ion adds one extra qubit and three additional vibrational modes. In this part of the Appendix, however, we show how the lesson learned from the previous calculations can be used to build a realistic and sophisticated crosstalk error model for imperfect 2-qubit MS gates.

The crucial point is that, as emphasised at the end of the last subsection,  the leading-order contribution of crosstalk comes from the extra unitary involving spectator qubits. This occurs due to crosstalk even for a perfect MS gate, which would not be subjected to off-resonant carriers, intensity and phase fluctuations, or Debye-Waller factors and residual qubit-phonon entanglement for the additional vibrational modes. Hence, to leading order, one may combine the microscopic error channel of the imperfect MS gate for  the active qubits~\cite{1810.09199}, and the above  coherent  effect of crosstalk~\eqref{eq:ct_error_coherent_unitary}.

The noisy MS gate on the active qubits can be represented by the ideal unitary, followed by a structured microscopic error channel different from depolarising noise. In particular, $\rho\to D_{MS}(U_{MS}\rho U_{MS}^\dagger)$, where $U_{MS}=\frac{1}{\sqrt{2}}(\id_4-\ii X_{i_1}X_{i_2})$ is the ideal MS gate between the pair of active ions $i_1,i_2\in G$; whereas $D_{MS}(\rho)$ admits a   Kraus decomposition~\eqref{kraus_decomp}  after retaining only  the most-likely operators that stem from the process-tomography numerical reconstruction of the microscopic quantum channel~\cite{1810.09199}, namely

\begin{equation}
\label{eq:kraus_ct_map}
    D_{MS}(\rho) \to \left\{ \begin{array}{ll}
         K_{0} =\sqrt{(1-p)}\quad  \id_4\\
        K_{1} =\sqrt{p \cdot 0.80} \quad  X_1X_2\\
        K_{2} =\sqrt{p \cdot  0.05} \quad Y_1\\
        K_{3} =\sqrt{p \cdot  0.05} \quad Y_2\\
        K_{4} =\sqrt{p \cdot  0.05} \quad X_1Z_2\\
        K_{5} =\sqrt{p \cdot  0.05} \quad Z_1X_2    \end{array} \right\}, 
\end{equation}

where $p$ is the MS gate error rate, which is one of the control variables in the simulations. 

Finally, for the spectator ions, the leading-order contribution stems  from the coherent application of the MS-gate unitary due to residual light.  To build a coherent model for this error source, we  consider that the residual couplings only have visible effects on the spectator qubits that are immediate neighbours of the active qubit pair (see the discussion in the main text). Accordingly, the leading-order coherent crosstalk can be modelled by  applying a $X_gX_n$ rotation of angle $\theta_c$  onto  every ion from the active qubit pair $g\in G$ and an neighbouring  spectator ion $n\in N$. This leads to the unitary

\begin{equation}
\label{eq:ct_error_coherent_unitary}
    R_{\rm ct}(\theta) 
    = \cos \frac{\theta_c(t)}{2} \id_4  -\ii \sin \frac{\theta_c(t)}{2} X_gX_n,
\end{equation}

This angle $\theta_c(t)$ is given by the fraction of the laser intensity that illuminates the neighbouring ion, and the angle $\Theta$ of the ideal MS gate,  e.g. $\Theta=\pi/2$ for a fully-entangling MS gate, leading to 

\begin{equation}
\theta_c (t)= \Theta \cdot \sqrt{\frac{I_{\rm r}(t)}{\langle I(t)\rangle}}.
\end{equation}

As discussed previously for single-qubit crosstalk~\eqref{eq:ct_angle_1q}, the residual intensity fluctuations could be inferred from experimental measures, or modelled stochastically. Moreover, one could also consider how the phase fluctuations give rise to slow drifts of the basis where the MS-gate is applied, which can also be estimated from experimental measurements. In this work, we will model  the effects of crosstalk in  coherent simulations as coherent $X_nX_g$ rotations of angle $\theta_c = \epsilon_{CT} \theta$ (Eq. \ref{eqCrosstalk}). This angle is fixed for a given crosstalk error rate and remains constant through the simulation.

As discussed in Eq.~\eqref{eq.coherent.crosstalk} of the main text, in the Clifford approximation, this crosstalk rotation of an angle $\theta_c$ corresponds to the application of  $X_gX_n$  with probability

\begin{equation}\label{eq.crosstalkRate}
    p_c = \sin^2\frac{\theta_c}{2}.
\end{equation}
 
Note that when the spectator ion is located between two active ions subjected to the MS gate, it will receive residual radiation from both addressed beams. This will double the residual amplitude of the field on the ion, and thus quadruple the corresponding intensity, changing the angle of crosstalk to twice the previous value

\begin{equation}
\theta_c' = \Theta \cdot \sqrt{\frac{4I_{\rm r}(t)}{\langle I(t)\rangle}} = 2\theta_C.
\end{equation}

Assuming that this crosstalk rotation angle is small, the effect on the probability of crosstalk will be a factor of 4, namely

\begin{equation}
\label{eq.crosstalkProbabilityIon}
    p_c' = \sin^2\left(\frac{\theta_c' }{2} \right)
    = 4p_c(1-p_c)\approx 4p_{c}.
    \end{equation}

In the numerical simulations, we use the parameter $ p_c$ to analyse the behaviour of crosstalk. From that parameter we obtain the angle of crosstalk as $\theta_c = 2 \arcsin\sqrt{p_c}$. If an ion is between two gate ions, we can use equation \eqref{eq.crosstalkProbabilityIon} to obtain the probability $p_c$ of those crosstalk events in the incoherent simulations, or we can multiply by 2 the angle of the crosstalk rotations in the wave-function simulations.

\subsection{Crosstalk noise for power-imbalanced two-qubit gates}\label{sec.appendixACstarkshift}

Let us now describe  the second, 
\textit{Stark-shift crosstalk error model} that arises from the following microscopic procedure to minimise the effect of crosstalk. As discussed previously, the state-dependent dipole force~\eqref{eq:state_dep_force} underlying the MS gate is obtained by combining a pair of bi-chromatic laser beams with opposite detunings that drive the red- and blue-sidebands of the centre-of-mass mode $\omega_{\rm COM}$ of the ion crystal

\begin{equation}
\label{eq:MS_res_conditions}
    \omega_{\rm L,1}=\omega_0+\delta, \hspace{1ex}     \omega_{\rm L,2}=\omega_0-\delta,\hspace{2ex} \delta\approx\omega_{\rm COM},
\end{equation}

where $\omega_0$ is the transition frequency between the two-qubit states. Additionally, in order to arrive to Eq.~\eqref{eq:state_dep_force}, both laser beams have the same intensity, leading to $\Omega_{1}=\Omega_{2}=:\Omega$. In case the intensities of the red- and blue-sidebands differ, one would obtain an additional differential  ac-Stark shift

\begin{equation}
    H_{\rm ac}=\sum_{i=1}^N\frac{\Omega_{\rm 1}^2-\Omega_{\rm 2}^2}{4|\delta|}\sigma_i^z+\sum_{j=1}^M\frac{\Omega_{\rm 1,j}^2-\Omega_{\rm 2,j}^2}{4|\delta|}\sigma_j^z,
\end{equation}

which would differ for active and spectator ions since they are subjected to different laser intensities and thus $\Omega_1\gg\Omega_{1,j},\Omega_2\gg\Omega_{2,j}$.
To cope with these shifts, one may modify the resonance conditions~\eqref{eq:MS_res_conditions} to

\begin{align}
    \omega_{\rm L,1}=&\omega_0+\frac{\Omega_{\rm 1}^2-\Omega_{\rm 2}^2}{2|\delta|}+\delta, \hspace{1ex} \nonumber \\    \omega_{\rm L,2}=&\omega_0+\frac{\Omega_{\rm 1}^2-\Omega_{\rm 2}^2}{2|\delta|}-\delta,
\label{eq:MS_res_conditions_power_imb}
\end{align}

to obtain again a state-dependent dipole force~\eqref{eq:state_dep_force}. However, note, that the spectator ions suffer a much smaller ac-Stark shift as they only see a small fraction of the laser light. Accordingly, their corresponding blue- and red-sidebands will no longer be nearly-resonant. As a consequence, the collective spin operator~\eqref{eq:coll_spin} becomes 

\begin{equation}
\tilde{J}_y(t)=\frac{1}{2}\sum_{i=1}^N\sigma_i^y+\frac{\ii}{2}\sum_{j=1}^M\frac{\Omega_j}{\Omega}\left(-\sigma_j^+\ee^{\ii\Delta t}+\sigma_j^{-}\ee^{-\ii\Delta t}\right)
\end{equation}
such that the contribution that depends on the spectator qubits rotates rapidly with a frequency $\Delta$ that scales with the difference of laser intensities seen by the active and spectator ions. Making sure that the power imbalance has the correct order of magnitude, these spectator terms will be far off-resonant, such that the MS gate will not entangle the spectator ions. 

According to this discussion, we can restrict the collective operator to the active qubits 

\begin{equation}
\tilde{J}_y(t)=\frac{1}{2}\sum_{i=1}^N\sigma_i^y,
\end{equation}

and model the effect on the off-resonant spectator ions via the residual ac-Stark shift

\begin{equation}
\label{eq:spectator_ac_shifts}
    H_{\rm ac}=\sum_{j=1}^M\frac{\Omega_{\rm 1,j}^2-\Omega_{\rm 2,j}^2}{4|\delta|}\sigma_j^z.
\end{equation}

Following a similar reasoning as in the above subsection, the presence of these crosstalk ac-Stark shifts and  the modified resonance conditions~\eqref{eq:MS_res_conditions_power_imb} can lead to additional error sources in the MS gate, e.g. intensity fluctuations can now lead to  residual qubit-phonon entanglement. However,  all these crosstalk effects will be proportional to the already-small MS-gate error via an additional small parameter that scales with  $(\omega_j\Omega)^2\ll 1$. Hence, to leading  order, we can again consider the imperfect MS gate on the pair of active qubits~\eqref{eq:kraus_ct_map}, whereas the spectator qubits evolve coherently under the residual ac-Stark shifts~\eqref{eq:spectator_ac_shifts}, the strength of which may vary due to intensity fluctuations. Note that these systematic AC Stark shifts can be determined and experimentally compensated for by, e.g.,~adjusting the phases of all subsequent gates on the spectator ions. 

On our numerical simulations, we fix the crosstalk error rate to be constant through the circuit simulation. We model these errors as in Eq.~(\ref{eq:crosstalk-starkshift-coherent}) for the coherent simulations, and as the Kraus map in Eq.~(\ref{eq:Kraus_incoherent_ac}) for the incoherent simulations.


\subsection{Crosstalk noise from scattered photons during the measurements}\label{appendix_measurement}

\begin{center}
    \begin{figure}
        \centering
        \includegraphics[width=0.95\columnwidth]{ 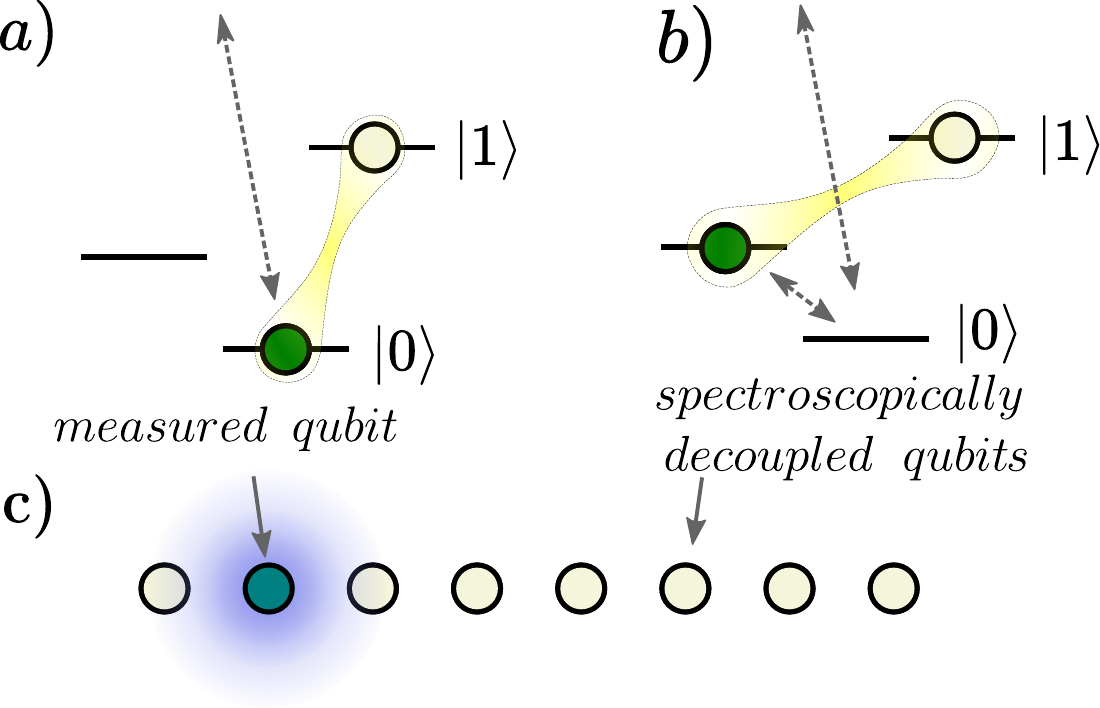}
        \caption{ \textbf{Qubit measurement and spectroscopic decoupling.}   \textbf{a)} During the measurement, a laser drives a transition from the $\ket{0}$ state to a metastable state with a fast decay. If the $\ket{0}$ state is measured, the ion scatters photons in all directions, potentially affecting the neighboring ions. \textbf{b)} To avoid this crosstalk noise, spectroscopic decoupling pulses can shift the population in the $\ket{0}$ state to a different energy level without a resonant transition at the frequency of the scattered photons. The original state can be recovered with another pulse sequence, preserving the original superposition. \textbf{c)} Sketch of the ion string. 
        }
        \label{fig.measurementcrosstalk}
    \end{figure}
\end{center}

During the measurement of an ion, we use a laser resonant with a transition that drives the $\ket{0}$ state to a metastable state. If the ion was in the $\ket{0}$ state, it will scatter photons from that transition in all directions, allowing us to detect it and differentiate the state from the \textit{dark} $\ket{1}$ state. The scattered photons from a measured $\ket{0}$ state can reach other ions in the ion string, potentially disturbing their quantum state. To avoid this problem, it is possible to apply spectroscopic decoupling pulses on the neighboring ions. During this process, the population in the $\ket{0}$ state is temporarily displaced to a different stable state from the ion energy levels using coherent pulses (See Fig. \ref{fig.measurementcrosstalk}).  In this way, the information stored in the qubit will not be affected by scattered photons from other measured ions. Therefore, we can neglect any effects from measurement crosstalk in our simulations. Further information on the topic can be found in \cite{nigg-science-345-302}.

\section{Details on the model for other error sources\label{sec:appendix_errormodels}}



 \subsection{Leakage and amplitude damping}

 The leakage process is a non-unital error channel that describes decay from the excited state to a \textit{leaked} state outside of the computational basis. This noise changes the gates and measurements applied to the qubit \cite{1810.09199}.
 
 For single-qubit gates, the lasers will only cause an off-resonant ac-Stark shift if the qubit is in the $\ket{L}$ state. As there is no coherence between the $\ket{L}$ state and the states in the computational basis, the effects of the single-qubit gates can be neglected. 
 
 For the two-qubit MS gate, we can consider two cases. If both of the qubits have leaked, the lasers will be highly off-resonant, and this will lead to ac-Stark shifts on the leaked qubits that can be neglected as in the single-qubit gate case.
 If only one of the qubits has leaked, the lasers will still be near-resonant with the sidebands of the unleaked qubit and its state will evolve realising a trajectory in phase space. During this trajectory, the spin of the qubit is entangled with the motional modes of the ion string, but the timing of the gate guarantees that the phase-space trajectory will be closed at the end of the pulse. To leading order, the evolution of this unleaked qubit is equivalent to the identity operator.
 
 As for the measurements, the result will be confused with the ground state $\ket{0}$, as the fluorescent measurement is not able to distinguish between $\ket{0}$ and $\ket{L}$ \cite{Schindler2013}.\\
 
The evolution of this system can  be described by the Kraus map with the following Kraus operators:
\begin{equation}
    \varepsilon_{AD} = \left\{ \begin{array}{ll}
         E_{A0} = \ket{0}\bra{0} +\ket{L}\bra{L} + \sqrt{1-\gamma} \ket{1}\bra{1}\\
        E_{A1} = \sqrt{\gamma_d}\ket{0}\bra{1} \\
        E_{A2} = \sqrt{\gamma_l}\ket{L}\bra{1}\end{array} \right.  
\end{equation}

where
\begin{equation}
    \gamma_d = \frac{\Gamma_d}{\Gamma}\left( 1-e^{-\Gamma t}\right), \quad   \gamma_l = \frac{\Gamma_l}{\Gamma}\left( 1-e^{-\Gamma t}\right),  
\end{equation}
and $\gamma = \gamma_d + \gamma_l$.\\

For the ion trap setup, $T_1 =1/\Gamma =  1.1s$ for $^{40}$Ca$^{+}$ ions, and $\frac{\Gamma_l}{\Gamma_d} = \frac{4}{9}$. In the numerical simulations, we use a Clifford approximation for this channel, as described in Eq.~(\ref{eq.leakageKraus}).

 These effects can be a problem, as they will change the effect of the circuit and affect the measurement results. In a QEC cycle in which we measure the stabilisers to obtain the syndrome, it can lead to an incorrect result that will affect the correction applied to the state. 
 
 To correct it, we need to apply first a leakage repumping sequence, in which we reset the qubit to the computational basis, and then we can  apply a round of QEC.

\subsection{Leakage repumping}
When a leakage error happens, and a qubit decays out of the code space to the leaked state, this transition is not detected. Therefore, we need to correct possible leakage errors that might occur on the qubits without measuring them, in a way that will to lowest order not affect the state of the qubit if there has not been any leakage event. 

\begin{center}
    \begin{figure}
        \centering
        \includegraphics[width=\columnwidth]{ 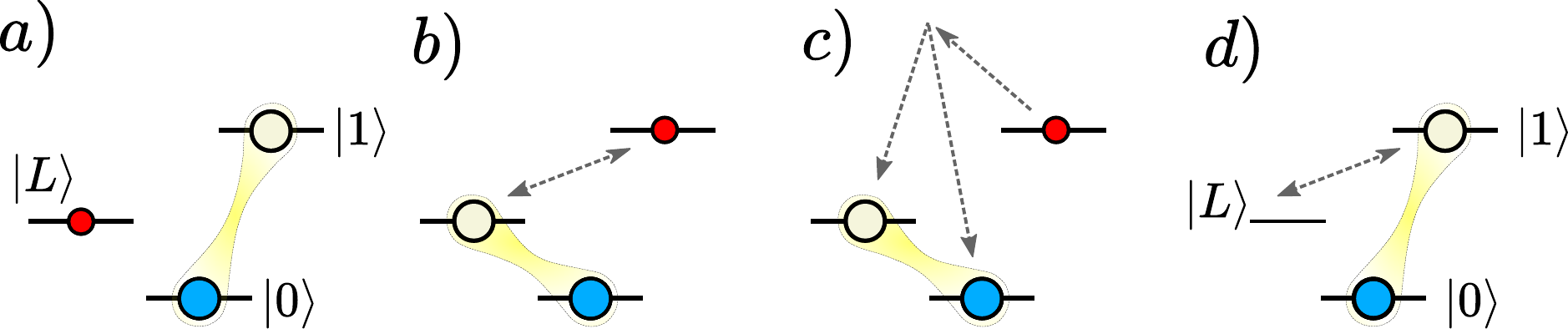}
        \caption{\textbf{Leakage repumping sequence: a)} initial state, where the initial populations are represented by the coloured circles. \textbf{b)} We apply a $\pi$ pulse between the states $\ket{1}$ and $\ket{L}$, swapping the populations. \textbf{c)} We apply a pulse between the $\ket{1}$ state and a metastable state, that will rapidly decay to either $\ket{L}$ or $\ket{0}$. \textbf{d)} A final $\pi$ pulse swaps the population in $\ket{L}$ to $\ket{1}$.  }
        \label{leakagerepump}
    \end{figure}
\end{center}

Using the leakage repumping sequence described in Fig.~\ref{leakagerepump}, we can bring the population in the leaked state back to the code space, while protecting the population and coherences in the code space if there was no leakage. This is achieved by first spectroscopically decoupling (\textit{hiding}) the population in the excited state by swapping it to the leaked state. Then, we apply a pulse that would bring any population in the leaked state (that is now occupying the excited state) to a metastable state that will rapidly decay to either the leaked state or the ground state.  Finally, the last pulse will bring the population in the leaked state, where the initial population in $\ket{1}$ was hidden, back to the excited state.

The total effect of this pulse sequence is equivalent to identity if the qubit was not leaked, as both the population and coherences remain protected. If the qubit had leaked, however, then it will be effectively reinitialised to either the ground or the excited state. In that case, the next round of QEC is expected to be able to correct that error.

\subsection{Leakage repumping errors.}

The leakage repumping sequence relies on single-qubit pulses that are susceptible to errors. We can do a simple estimate of the errors that can happen during the process and the final result on the qubit by checking the effects of single errors in the pulses. To this end, we begin in a state that is either on the leaked state $\ket{L}$ or in the computational subspace. Then, we assume an error in one of the three single-qubit gates involved in the sequence, which are events that will appear with probability $p_{sg}$.

Let us follow the case in which the initial state was in the $\ket{L}$ state (red circle in figure \ref{leakagerepump}). 
If an error happened on the first gate, some of the population in the leaked state might remain leaked during the rest of the process. Following the evolution of that population, we can see that the following pulses will still bring the population back to the code space. In this case, that error event does not lead to an error, and the leakage repumping sequence works. If an error happens on the second gate, some population in the $\ket{1}$ state will remain there, and the last pulse will bring that population to the $\ket{L}$ leaked state. This process will happen with probability $p_{sg}$ and is equivalent to leakage.
We obtain the same result if the error happens in the last pulse of the sequence, in which the population in the $\ket{L}$ leaked state will remain leaked.

Let us examine now the case in which the initial state was in the computational subspace. If the error occurs in the first pulse, some of the population that was initially in the $\ket{1}$ state will end up being reassigned at random with probability $\frac{1}{2}$ to either $\ket{0}$ or $\ket{1}$ in the second pulse. This event will occur with probability $p_{sg}$, and we consider the result as dephasing (as the decay will destroy the coherences in the state) with probability $p_{sg}/2$ or amplitude damping with $p_{sg}/2$. If we place the error in the second pulse, there will be no consequences as the initial population in the leaked state was zero. However, an error in the last pulse can lead to leakage with probability $p_{sg}$.

This can be summarised in:
\begin{enumerate}
    \item If the qubit was leaked: we have a probability of twice the single gate error probability $2p_{sg}$ to bring the qubit back to the leaked state.
    \item If the qubit was not leaked: then, with probability $p_{sg}$ it can end in the leaked state, with $p_{sg}/2$ it can suffer amplitude damping and with $p_{sg}/2$ it can suffer dephasing.
\end{enumerate}

\section{Details on the numerical simulations.}

In the numerical simulations used to generate the plots shown in the paper, we used both coherent and incoherent simulations, as indicated on each figure.
As stated in the main text, we use the open-source software PECOS  \cite{1812.04735,PECOSGit} as the platform to run the simulations. Incoherent simulations are implemented using the stabiliser simulator included in PECOS, while the coherent simulations are implemented using the state-vector simulator from ProjectQ  \cite{1612.08091,1604.01401} as a backend to PECOS.\\

To compute the logical error rates in Figs.~\ref{Fig:ThreeChecksNoIgor}, \ref{Fig: refocussingResults}, and \ref{Fig:starkshift}, we use standard Monte Carlo sampling: we generate an error configuration according to the error model, and run the QEC round to obtain a final result of successful recovery or failure event. The repetition of this process allows us to obtain an estimate of the logical error probability for a given set of error parameters. 

To obtain the error bar of each estimate, we use the error from the binomial distribution $err = \sqrt{p_{log}(1-p_{log})/n}$.  The dependence of the error on the logical error rate $p_{log}$ implies that a larger number of samples is required for the low $p_{log}$ regime. Therefore, we use a different number of samples for each point in the figures, ranging from $10^5$ to $10^7$ samples per point, until the error bars are smaller than the markers used in the plots.  

The error model from the simulations of Fig.~\ref{Fig:singleflag_cvsi_onlyc} contains only crosstalk events. For the coherent simulations, this implies the  application of additional coherent rotations after every MS gate. These events are not random, but a systematic error. Therefore, the only stochastic process in this simulation is the process of measuring, where the wave function is partially collapsed. Since the total number measurement events for the worst case scenario is only 12 (measuring the 6 stabilisers twice), it is possible to simulate individually each measurement outcome path and compute analytically from the wave function the resulting logical error probability. To compute the total logical error probability, we can sum the result of each \textit{path}, weighted by the probability of that measurement combination:

\begin{equation}
    p_{Logical} = \sum_{j}^{2^{12}} \prod_i m_i^{(j)} p_L(\{m_i\})_j)
\end{equation}
Here, $m_i$ is the outcome of a measurement in the $i^{th}$ measurement,  $\{m_i\}_j$ is a combination of measurement outcomes (or measurement path) and $p_L(\{m_i\})_j)$ is the logical error probability for the resulting state after path $j$. Following this alternative, exhaustive sampling of all measurement outcomes, we can accurately estimate the logical error probability for a given crosstalk value using only $2^{12}$ simulations, independently of the crosstalk error amplitude.

\end{document}